\documentclass[11pt,a4paper]{article}
\usepackage{jcappub_nt}
\usepackage[T1]{fontenc}
\usepackage{babel}
\usepackage{array}
\usepackage{float}
\usepackage{mathtools}
\usepackage{multirow}
\usepackage{stmaryrd}
\usepackage{wasysym}
\usepackage{bm}

\title{Spatially covariant gravity with nonmetricity}

\author{Yang Yu,}
\author{Zheng Chen}
\author[1]{and Xian Gao\note{Corresponding author.}}
\affiliation{School of Physics and Astronomy, Sun Yat-sen University, Zhuhai 519082, China}
\emailAdd{yuyang69@mail2.sysu.edu.cn}
\emailAdd{chenzh396@mail2.sysu.edu.cn}
\emailAdd{gaoxian@mail.sysu.edu.cn}

\abstract{
Scalar fields play an important role in constructing modified gravity theories. In the case of a single scalar field with timelike gradient, the corresponding Lagrangian in the unitary gauge takes the form of spatially covariant gravity (SCG), which is proved useful in analyzing and extending the generally covariant theories. In this work, we apply the SCG method to the scalar-nonmetricity theory, of which the Lagrangian is built of the nonmetricity tensor and a scalar field. We derive the 3+1 decomposition of the geometric quantities and especially  covariant derivatives of the scalar field up to the third order in the presence of a nonvanishing nonmetricity tensor. By fixing the unitary gauge, the resulting Lagrangian takes the form of a SCG with nonmetricity, in which all the ingredients are spatial tensors. We then exhaust the scalar monomials of SCG with nonmetricity up to $d=3$ with $d$ the total number of derivatives. Since the disformation tensor plays as an auxiliary variable, we take the Lagrangian with $d=2$ as an example to show that after solving the disformation tensor, we can get an effective SCG theory for the metric variables but with modified coefficients. Our results provides a novel approach to extending the scalar-nonmetricity theory.
}

\begin{document}
	\maketitle

\section{Introduction}

Theories of modified gravity typically involve extra degrees of freedom
in addition to the two tensorial degrees of freedom of General Relativity.
The simplest and most efficient approach is to introduce extra scalar field(s). A successful example is the construction of higher derivative scalar-tensor theory with a single scalar degree of freedom, which includes the rediscovery of Horndeski theory \cite{Horndeski:1974wa,Nicolis:2008in,Deffayet:2011gz,Kobayashi:2011nu}, as well as the development of degenerate higher-order scalar-tensor theory \cite{Langlois:2015cwa,Crisostomi:2016czh,BenAchour:2016fzp}.

An alternative approach to introducing scalar degree(s) of freedom is to break the full spacetime diffeomorphism, particularly by breaking temporal symmetry while retaining only spatial diffeomorphism.
This idea can be traced back to ghost condensation \cite{Arkani-Hamed:2003pdi}
and was extensively developed in the effective field theory of inflation \cite{Cheung:2007st,Gubitosi:2012hu,Ashoorioon:2018uey} and Ho\v{r}ava gravity \cite{Horava:2009uw}. In \cite{Gao:2014soa,Gao:2014fra}, a class of spatially covariant gravity (SCG) theories that propagate a single scalar degree of freedom was constructed, which was further extended by introducing the velocity
of lapse function \cite{Gao:2018znj,Gao:2019lpz}. 
A large class of SCG Lagrangians without any scalar degree of freedom has been investigated in \cite{Gao:2019twq,Hu:2021yaq} and applied in the study of cosmology and black holes \cite{Iyonaga:2021yfv,Hiramatsu:2022ahs,Saito:2023bhn,Chakraborty:2023jek}\footnote{Such kind of theories has also been discussed in the context of Ho\v{r}ava gravity \cite{Zhu:2011xe,Zhu:2011yu}. See also extensions with a dynamical lapse function \cite{Lin:2020nro} as well as with auxiliary constraints \cite{Yao:2020tur,Yao:2023qjd}.}.
Additionally, constraints from gravitational waves on SCG have
been explored in \cite{Gao:2019liu,Zhu:2022dfq,Zhu:2022uoq,Zhu:2023wci},
from which one can see that SCG provides a broad framework to study
various modified gravity theories in a unifying manner.

In the case of a single scalar field, generally covariant scalar-tensor theory (GST) and SCG have a one-to-one correspondence with each other.
Indeed, when the gradient of the scalar field is timelike, it is possible
to choose the time coordinate to be a function of the scalar field, which is dubbed the unitary gauge in the literature. Since the scalar field is chosen to be the time coordinate, the GST theory appears to be a theory with only variables from the gravity side respecting spatial diffeomorphism, i.e., an SCG theory. 
Conversely, a SCG theory can be recast as a GST theory by the gauge recovering procedure (St\"{u}ckelberg trick). Such a correspondence between SCG and GST theories has been discussed in detail in \cite{Gao:2020qxy,Gao:2020yzr,Hu:2021bbo} (see also \cite{Joshi:2021azw,Joshi:2023otx}). 

In recent years, gravity theories based on non-Riemannian geometry \cite{Weyl:1918ib,Cartan:1923zea,Hehl:1994ue}, which assume nonvanishing torsion and/or nonmetricity tensors, have attracted much attention. 
Generally, the affine connection $\varGamma_{\phantom{c}ab}^{c}$ is treated as an independent variable, which extends the geometric description of gravitation. The theory with a general affine connection is thus dubbed metric-affine gravity (MAG), which has been
studied extensively \cite{Sotiriou:2006qn,Capozziello:2009mq,Vitagliano:2010sr,Iosifidis:2018diy,Iosifidis:2018zwo,Iosifidis:2018jwu,BeltranJimenez:2019acz,Percacci:2020ddy,BeltranJimenez:2020sqf,Cabral:2020fax,Iosifidis:2021tvx,Iosifidis:2021fnq,Iosifidis:2021bad,Baldazzi:2021kaf,Pradisi:2022nmh,Capozziello:2022zzh}.
By coupling MAG to scalar field(s), one gets the so-called ``metric-affine scalar-tensor theory'' \cite{Aoki:2018lwx,Helpin:2019kcq,Helpin:2019vrv,Aoki:2019rvi,Kubota:2020ehu},
which can be seen as an analogue of the scalar-tensor theory in Riemannian
geometry. 
In \cite{Dong:2021jtd,Jimenez-Cano:2022arz,Dong:2022cvf}, propagation of gravitational waves has been discussed, which gives constraints on the relevant theory. MAG also generates ``gauge theories of gravity'' by applying the Yang-Mills theory to gravity (see \cite{Blagojevic:2012bc} for a review).

One subclass of the general MAG that is extensively studied is teleparallel gravity (TG) (see \cite{Bahamonde:2021gfp} for a review), which assumes absolute teleparallelism, i.e., vanishing curvature.
By assuming further a vanishing torsion, one gets the so-called symmetric teleparallel gravity (STG), which is based on the nonmetricity tensor and is proved to be equivalent to general relativity up to a boundary term \cite{Nester:1998mp}. STG has been extensively studied in the literature \cite{Nester:1998mp,Adak:2005cd,Adak:2006rx,Adak:2008gd,Mol:2014ooa,BeltranJimenez:2017tkd,Jarv:2018bgs,Runkla:2018xrv,Hohmann:2018wxu,Adak:2018vzk,Lu:2019hra,BeltranJimenez:2019tme,Lazkoz:2019sjl,Koivisto:2019ggr,Xu:2020yeg,DAmbrosio:2020nqu,Flathmann:2020zyj,Zhao:2021zab,BeltranJimenez:2022azb,Albuquerque:2022eac,Dimakis:2022rkd,Chen:2022wtz,Zhang:2023scq,Hu:2023ndc}.
Similarly, with torsion tensor only, the (torsional) teleparallel
equivalent of general relativity \cite{Maluf:2013gaa} is also studied in \cite{Ferraro:2006jd,Chen:2010va,Zheng:2010am,Geng:2011aj,Bamba:2013jqa,Kofinas:2014owa,Li:2018ixg,Ualikhanova:2019ygl,Li:2020xjt,Bose:2020xdz,Li:2021wij,Li:2022vtn,Zhang:2022xmm,Li:2023fto,Feng:2023veu,Hu:2023juh,Rao:2023doc,Bajardi:2023gkd,Capozziello:2023giq}.

One lesson from the study of the SCG correspondence of GST is that the former has much simpler form and in particular, it is more straightforward to extend the theory in the SCG form instead of the original GST form.
Can we take advantage of SCG in studying MAG coupled to a scalar field? This work is devoted to applying this idea to the scalar-nonmetricity theory.

To this end, we first need to identify the basic building blocks in the unitary gauge of a general scalar-nonmetricity theory. 
On the gravity side, besides the curvature quantities, the basic variables are the decomposition of the nonmetricity tensor (or equivalently the disformation tensor) on the spatial hypersurfaces. On the scalar field side, the essential quantities are the decompositions of the generally covariant derivatives of the scalar field $\nabla\phi,\nabla\nabla\phi,\nabla\nabla\nabla\phi,\dots$
parallel and orthogonal to the spatial hypersurfaces. In the unitary
gauge, since all the decomposed quantities are spatial tensors, we
can use them as the basic building blocks to construct the general SCG with metric and nonmetricity variables.
In this work, we concentrate only on the polynomial type Lagrangians,
and exhaust and classify all the scalar monomials up to $d=3$ with
$d$ the total number of derivatives.

In this work, we assume the nonmetricity tensor and equivalently the
disformation tensor to be auxiliary variables with no temporal derivative. In principle, we may
solve the nonmetricity tensor in terms of the metric variables. The
resulting Lagrangian is nothing but a subclass of SCG with pure metric
variables. In
this work, we employ the polynomial type Lagrangian with $d=2$ as
an example. 

This paper is organized as follows. In Sec. \ref{sec:pre}, we analyze
the basic building blocks by making the 3+1 decomposition without
the metric compatible condition. In Sec. \ref{sec:mono}, we build
and exhaust all the scalar monomials up to $d=3$. In Sec. \ref{sec:quad},
we take Lagrangian with $d=2$ as an example and show the effective
SCG Lagrangian after solving the disformation tensor. In Sec. \ref{sec:con},
we summarize our discussion. Throughout this paper we choose the unit
$8\pi G=1$ and the convention for the metric $\{-,+,+,+\}$. The
indices $\{a,b,c,\dots\}$ stand for the 4-dimension covariant coordinates
and the indices $\{i,j,k,\cdots\}$ stand for spatial coordinates, respectively.

\section{Preliminary} \label{sec:pre}

In this section, we briefly review some basic concepts concerning the nonmetricity and derive the 3+1 decomposition of derivatives of the scalar fields.

\subsection{Geometry}

In Riemannian geometry, the connection is assumed to be metric-compatible and torsionless,  thus uniquely determined by the metric, i.e., the Levi-Civita connection. 
In this work, we consider the non-Riemannian geometry, where the affine connection is assumed to be independent of the metric. Such a theory is dubbed metric-affine gravity (MAG) in the literature.

In our work, we consider a subclass of MAG theory in which the connection is not metric-compatible but is still free of torsion, i.e.,
$T_{\phantom{c}ba}^{c}\coloneqq\varGamma_{\phantom{c}ab}^{c}-\varGamma_{\phantom{c}ba}^{c}=0$.
The nonmetricity tensor is defined as 
\begin{align}
Q_{cab} & \coloneqq\nabla_{c}g_{ab}=\partial_{c}g_{ab}-\varGamma_{\phantom{d}ac}^{d}g_{db}-\varGamma_{\phantom{d}bc}^{d}g_{ad},
\end{align}
where $g_{ab}$ is the spacetime metric and $\varGamma_{\phantom{c}ab}^{c}$
is the affine connection. Once we consider such torsionless non-Riemannian
geometry, the difference between the affine connection $\varGamma_{\phantom{c}ba}^{c}$
and the metric-compatible Levi-Civita connection $\mathring{\varGamma}_{\phantom{c}ba}^{c}$
is encoded in the disformation tensor,
\begin{align}
L_{\phantom{c}ba}^{c} & \coloneqq\varGamma_{\phantom{c}ba}^{c}-\mathring{\varGamma}_{\phantom{c}ba}^{c}.
\end{align}
Throughout this paper, an overcircle ``$\circ$'' denotes quantities
adapted to the Levi-Civita connection. The disformation tensor can
be written as a function of the nonmetricity tensor
\begin{align}
L_{cab} & =\frac{1}{2}\left(Q_{cab}-Q_{abc}-Q_{bca}\right),
\end{align}
or inversely we have
\begin{align}
Q_{cab} & =-L_{bac}-L_{abc}.\label{eq:2.4}
\end{align}
Note that the disformation tensor is symmetric with respect to its last two
indices, i.e., $L_{abc}\equiv L_{acb}$.

In this work, we will apply the method of spatially covariant gravity to the scalar-nonmetricity theory. To this end, we have to introduce a proper foliation
structure of the spacetime and decompose all geometric quantities,
including the metric, nonmetricity tensor as well as derivatives of
the scalar field, into their temporal and spatial parts. As usual,
we choose $n^{a}$ as the normal vector to the spatial hypersurfaces,
which is normalized and timelike, i.e., $n^{a}n_{a}=-1$. The induced
metric on the hypersurface is defined by 
\begin{equation}
h_{ab}=g_{ab}+n_{a}n_{b}.
\end{equation}
The acceleration and the extrinsic curvature are defined as usual
by the Lie derivatives
\begin{equation}
a_{a}=\pounds_{\bm{n}}n_{a},\qquad K_{ab}=\frac{1}{2}\pounds_{\bm{n}}h_{ab},\label{acce_extcur}
\end{equation}
where $\pounds_{\bm{n}}$ is the Lie derivative with respect to $n^{a}$.
We emphasize that since the Lie derivative is irrelevant to any specific
connection, $a_{a}$ and $K_{ab}$ defined in (\ref{acce_extcur})
in terms of Lie derivatives are thus universal, which in particular,
are exactly the same as in the Levi-Civita case.

With the normal vector $n^{a}$ and the induced metric $h_{ab}$,
any spacetime tensor can be decomposed into the temporal and spatial
parts. In particular, the disformation tensor is decomposed as
\begin{eqnarray}
L_{abc} & = & -n_{a}n_{b}n_{c}L_{\bm{n}\bm{n}\bm{n}}+n_{a}n_{b}L_{\bm{n}\bm{n}\hat{c}}+n_{a}n_{c}L_{\bm{n}\bm{n}\hat{b}}+n_{b}n_{c}L_{\hat{a}\bm{n}\bm{n}}\nonumber \\
 &  & -n_{a}L_{\bm{n}\hat{b}\hat{c}}-n_{b}L_{\hat{a}\hat{c}\bm{n}}-n_{c}L_{\hat{a}\hat{b}\bm{n}}+L_{\hat{a}\hat{b}\hat{c}}.\label{dec_Labc}
\end{eqnarray}
Throughout this paper we follow the notation in \cite{Deruelle:2009zk}
such that an index replaced by ``$\bm{n}$'' denotes contraction
with the normal vector, and an index with a hat denotes projection
by the induced metric, e.g., $L_{\bm{n}\bm{n}\hat{a}}\equiv n^{b}n^{c}h_{a}^{a'}L_{bca'}$
etc. For the sake of simplicity, we define a set of six spatial tensors
\begin{align}
L^{(1)} & \coloneqq L_{\bm{n}\bm{n}\bm{n}}, & L_{a}^{(2)} & \coloneqq L_{\bm{n}\bm{n}\hat{a}}, & L_{a}^{(3)} & \coloneqq L_{\hat{a}\bm{n}\bm{n}},\nonumber \\
L_{ab}^{(4)} & \coloneqq L_{\hat{a}\hat{b}\bm{n}}, & L_{ab}^{(5)} & \coloneqq L_{\bm{n}\hat{a}\hat{b}}, & L_{cab}^{(6)} & \coloneqq L_{\hat{c}\hat{a}\hat{b}},\label{projdis}
\end{align}
which are the independent projections of the disformation tensor on
the spatial hypersurface. Similarly, we can also define the projected
tensors for the nonmetricity tensors, which can be related by $L^{(1)},\cdots,L^{(6)}$
through the relation (\ref{eq:2.4}).

Since the affine connection is torsion-free, the spatial curvature tensor is defined by
\begin{align}
^{3}\!R_{\phantom{b}acd}^{b}A_{b} & \equiv\mathrm{D}_{c}\mathrm{D}_{d}A_{a}-\mathrm{D}_{d}\mathrm{D}_{c}A_{a},\label{RieTen3d}
\end{align}
where $A_{a}$ is an arbitrary spatial tensor, and $\mathrm{D}_{a}$ is the spatial derivative defined by (e.g.) $\mathrm{D}_{a} A_{b} = h_{a}^{a'} h_{b}^{b'} \nabla_{a'} A_{b'}$. Please note $^{3}\!R_{abcd}$
is different from $R_{\hat{a}\hat{b}\hat{c}\hat{d}}$ or other decompositions
of the spacetime curvature tensor $R_{abcd}$. In particular, the spacetime
curvature tensor $R_{abcd}$ has different symmetry from that of $\mathring{R}_{abcd}$,
which is the Riemann tensor constructed by the Levi-Civita connection,
i.e.,
\begin{equation}
^{3}\!\mathring{R}_{\phantom{b}acd}^{b}A_{b}\equiv\mathring{\mathrm{D}}_{c}\mathring{\mathrm{D}}_{d}A_{a}-\mathring{\mathrm{D}}_{d}\mathring{\mathrm{D}}_{c}A_{a},\label{RieTen3d_LC}
\end{equation}
with $\mathring{\mathrm{D}}_{a}$  the spatial derivative compatible with the spatial metric $h_{ab}$.
We define the spatial Ricci tensors to be
\begin{equation}
^{3}\!R_{ab}\coloneqq h^{cd}\,{}^{3}\!R_{cadb}\equiv h^{cd}R_{\hat{c}\hat{a}\hat{d}\hat{b}}+2K_{a[c}K_{b]}^{c},\label{RicTen3d}
\end{equation}
and similarly 
\begin{equation}
^{3}\!\mathring{R}_{ab}\coloneqq h^{cd}\,{}^{3}\!\mathring{R}_{cadb}.\label{RicTen3d_LC}
\end{equation}
Note generally $^{3}\!R_{ab}$ is not symmetric anymore, which is
related to $^{3}\!\mathring{R}_{ab}$ by (\ref{eq:B.3}) (in terms
of spatial tensors including the disformation tensors and spatial
derivatives of them). At this point, note with a non-trivial connection,
there could be some alternative contractions, e.g. $h^{cd}R_{\hat{a}\hat{c}\hat{d}\hat{b}}$,
which we do not consider in this work\footnote{In principle such more general contractions of the curvature tensor
can be introduced.}. 

By definition, the extrinsic curvature is the Lie derivative of the
spatial metric, which has nothing to do with any specific connection.
In fact we have the following relation
\begin{equation}
K_{ab}=\frac{1}{2N}\left(\tilde{\pounds}_{\bm{t}}h_{ab}-\mathring{\mathrm{D}}_{a}N_{b}-\mathring{\mathrm{D}}_{b}N_{a}\right)=\frac{1}{2N}\left(\tilde{\pounds}_{\bm{t}}h_{ab}-\mathrm{D}_{a}N_{b}-\mathrm{D}_{b}N_{a}-2N^{c}L_{c\hat{a}\hat{b}}\right),
\end{equation}
where $\tilde{\pounds}_{\bm{t}}h_{ab}\equiv h_{a}^{a'}h_{b}^{b'}\pounds_{\bm{t}}h_{a'b'}$
with $t^{a}$ the time flow vector, and $N^{a}$ is the shift vector.

\subsection{Decomposition of derivatives of the scalar field}

When considering metric-affine gravity coupled with a scalar field,
we must also decompose the derivatives of the scalar field. In
the case of a Levi-Civita connection, the relevant decomposition
can be found in \cite{Gao:2020yzr}. Here, we perform
the decomposition in the presence of a nonvanishing nonmetricity tensor.
We emphasize that this decomposition is performed with respect
to an arbitrary foliation with a timelike normal vector $n^{a}$.
In particular, we have not fixed any specific coordinates. 

For the first-order derivative of the scalar field, we have
\begin{align}
\nabla_{a}\phi & =-n_{a}\pounds_{\bm{n}}\phi+\text{D}_{a}\phi,\label{cd1sca_dec}
\end{align}
where $\text{D}_{a}$ is the projected derivative defined by $\text{D}_{a}\phi\coloneqq h_{a}^{b}\nabla_{b}\phi$.
Clearly, (\ref{cd1sca_dec}) is the same as that in the metric theory.
For the second-order derivative of the scalar field, we have
\begin{equation}
\nabla_{a}\nabla_{b}\phi  =n_{a}n_{b}A-2n_{(a}B_{b)}+\varDelta_{ab},\label{cd2sca_dec}
\end{equation}
with
\begin{align}
A & =\pounds_{\bm{n}}^{2}\phi-a_{a}\mathrm{D}^{a}\phi+L^{(1)}\pounds_{\bm{n}}\phi-L_{a}^{(3)}\mathrm{D}^{a}\phi,\label{cd2sca_dec_A}\\
B_{b} & =-\left(a_{b}-L_{b}^{(2)}\right)\pounds_{\bm{n}}\phi+\pounds_{\bm{n}}\mathrm{D}_{b}\phi-\left(K_{ab}+L_{ab}^{(4)}\right)\mathrm{D}^{a}\phi,\\
\varDelta_{ab} & =-\left(K_{ab}-L_{ab}^{(5)}\right)\pounds_{\bm{n}}\phi+\text{D}_{a}\text{D}_{b}\phi,\label{cd2sca_dec_Dlt}
\end{align}
where $B_{b}$ and $\varDelta_{ab}$ are spatial tensors. At this
point, note that by definition
\begin{equation}
\mathrm{D}_{a}\phi\equiv\mathring{\mathrm{D}}_{a}\phi\equiv h_{a}^{b}\partial_{b}\phi, \label{D1phi}
\end{equation}
which has nothing to do with the connection and thus the nonmetricity
tensor. On the other hand, the second-order derivatives $\text{D}_{a}\text{D}_{b}\phi$
implicitly include the nonmetricity tensor. By making use of the
relation
\begin{equation}
\nabla_{a}\nabla_{b}\phi=\mathring{\nabla}_{a}\mathring{\nabla}_{b}\phi-L_{\phantom{c}ba}^{c}\nabla_{c}\phi,\label{DDphi_rel}
\end{equation}
as well as (\ref{D1phi}), we find
\begin{equation}
\mathrm{D}_{a}\mathrm{D}_{b}\phi=\mathring{\mathrm{D}}_{a}\mathring{\mathrm{D}}_{b}\phi -L_{cab}^{(6)}\text{D}^{c}\phi,\label{DDphi_rel_xpl}
\end{equation}
with $\mathring{\mathrm{D}}_{a}$ being the spatial derivative compatible
with the induced metric $h_{ab}$. As a result, we can also replace
$\mathrm{D}_{a}\mathrm{D}_{b}\phi$ by $\mathring{\mathrm{D}}_{a}\mathring{\mathrm{D}}_{b}\phi$
with an additional term involving $L_{cab}^{(6)}$.

For completeness, we also show the decomposition of the third-order derivative, given by 
\begin{equation}
\nabla_{c}\nabla_{a}\nabla_{b}\phi  =-n_{c}n_{a}n_{b}U+2n_{c}n_{(a}V_{b)}+n_{a}n_{b}W_{c}-n_{c}X_{ab}-2Y_{c(a}n_{b)}+Z_{cab},\label{cd3sca_dec}
\end{equation}
with
\begin{align}
U & =\pounds_{\bm{n}}A-2a^{a}B_{a}+2L^{(1)}A-2L_{a}^{(3)}B^{a},\\
V_{b} & =-a_{b}A+L_{b}^{(2)}A+\pounds_{\bm{n}}B_{b}-K_{b}^{\phantom{b}a}B_{a}+L^{(1)}B_{b}-L_{db}^{(4)}B^{d}-a^{d}\varDelta_{db}-L_{d}^{(3)}\varDelta_{\phantom{d}b}^{d},\\
W_{c} & =\text{D}_{c}A+2L_{c}^{(2)}A-2K_{c}^{\phantom{c}d}B_{d}-2L_{dc}^{(4)}B^{d},\\
X_{ab} & =-2a_{(a}B_{b)}+2L_{(a}^{(2)}B_{b)}+\pounds_{\bm{n}}\varDelta_{ab}-2K_{(a}^{\phantom{(a}d}\varDelta_{b)d}-2L_{d(a}^{(4)}\varDelta_{\phantom{d}b)}^{d},\\
Y_{ca} & =-K_{ca}A+L_{ac}^{(5)}A+\text{D}_{c}B_{a}+L_{c}^{(2)}B_{a}-K_{c}^{\phantom{c}d}\varDelta_{ad}-L_{dc}^{(4)}\varDelta_{\phantom{d}a}^{d},\\
Z_{cab} & =-2K_{c(a}B_{b)}+2L_{c(a}^{(5)}B_{b)}+\text{D}_{c}\varDelta_{ab}.
\end{align}
By plugging (\ref{cd2sca_dec_A})-(\ref{cd2sca_dec_Dlt}) into the
above, we can obtain the explicit expressions for the coefficients $U$,
$V_{b}$, etc., which are shown in Appendix \ref{app:cd3phi}.

It is interesting to note that when expressed in terms of $\mathrm{D}_{a}$, $L_{cab}^{(6)}$ does not manifestly arise in the decomposition of
derivatives of the scalar field. However, $L_{cab}^{(6)}$ is actually
present implicitly in the spatial derivative $\mathrm{D}_{a}$, which
manifests itself explicitly by splitting $\mathrm{D}_{a}$ into $\mathring{\mathrm{D}}_{a}$
and additional nonmetricity terms. This can be seen transparently
in (\ref{DDphi_rel_xpl}) and similarly for other terms involving
$\mathrm{D}_{a}$.

As a consistency check, when the nonmetricity tensor is vanishing
(or equivalently by setting $L_{cab}=0$), all the decompositions above reduce to those in the case of the Levi-Civita connection
\cite{Gao:2020yzr}. It is also interesting to note that, up to the
second-order derivative of the scalar field, the Lie derivative of
the disformation tensor $\pounds_{\bm{n}}L$ does not arise. As a
result, the nonmetricity tensor (or equivalently, the disformation
tensor) is nondynamical and acts as an auxiliary field up to the
second order in derivatives of the scalar field. However, $\pounds_{\bm{n}}L$
will arise in the decomposition for the third derivative of the scalar
field, which implies that if we consider metric-affine gravity
coupled to a scalar field with third-order derivatives, the number
of degrees of freedom will drastically increase compared to the
case up to the second order in derivatives. 

\subsection{Unitary gauge} \label{subsec:unig}

One of the original motivations for considering spatially covariant gravity is its equivalence to the scalar-tensor theory when the scalar field $\phi$ possesses a timelike gradient. In this case, when performing the 3+1 decomposition, the spacelike hypersurfaces can be chosen as the $\phi=\mathrm{const.}$ hypersurfaces themselves.
The same procedure can also be applied when considering a general metric-affine gravity coupled to a scalar field. 

Precisely, we choose the normal vector field $n_{a}$ used for the 3+1 decomposition to be the gradient of the scalar field itself, i.e.,
\begin{align}
n_{a}\to u_{a} & \equiv-\frac{\nabla_{a}\phi}{\sqrt{2X}}\qquad\text{with}\quad X\coloneqq-\frac{1}{2}\nabla_{a}\phi\nabla^{a}\phi.
\end{align}
This corresponds to the so-called ``unitary gauge'' in the literature\footnote{Some authors define the ``unitary gauge'' as a specific choice of time coordinate, i.e., fixing $t=\phi$. Although this can always be done, it is actually unnecessary. We emphasize that we refer to the ``unitary gauge'' merely as choosing a specific foliation or, equivalently, fixing the normal vector $n_{a}$, which itself has nothing to do with any specific coordinate. The advantage of defining the unitary gauge in this way is that all the expressions of decomposition can be written in a generally covariant manner.}. 
In the unitary gauge, since the value of the scalar field is constant on each hypersurface, all the spatial derivatives of the scalar field vanish
\begin{equation}
\overset{\text{u}}{\mathrm{D}}_{a}\phi  \equiv\overset{\text{u}}{h}{}_{a}^{a'}\nabla_{a'}\phi=0\qquad\text{with}\quad\overset{\text{u}}{h}{}_{a}^{a'}=g_{a}^{a'}+u_{a}u^{a'},
\end{equation}
where a hat ``u'' stands for quantities defined in the ``unitary
gauge'', i.e., with $u_{a}\propto\nabla_{a}\phi$. With this setting,
the expressions of decomposition in the previous subsection are
significantly simplified. In the rest of this paper, since we always
work in the unitary gauge, we will omit the hat ``u'' for simplicity.

As an illustration of our formalism, let us consider the Horndeski
Lagrangian $\mathcal{L}_{4}^{\text{H}}$ \cite{Horndeski:1974wa,Nicolis:2008in,Deffayet:2011gz,Kobayashi:2011nu},
\begin{equation}
\mathcal{L}_{4}^{\text{H}}=G_{4}\left(\phi,X\right){}^{4}\!R+G_{4,X}\left(\phi,X\right)\left[\left(\boxempty\phi\right)^{2}-\phi^{ab}\phi_{ab}\right].\label{Lag4}
\end{equation}
We will make use of the relations
\begin{equation}
^{4}\!R={}^{4}\!\mathring{R}+L_{\phantom{a}ba}^{a}L_{\phantom{bc}c}^{bc}-L_{\phantom{c}bc}^{a}L_{\phantom{db}a}^{bc}+\mathring{\nabla}_{c}\left(L_{\phantom{cb}b}^{cb}-L_{\phantom{bc}b}^{bc}\right),
\end{equation}
and (\ref{DDphi_rel}), where $L_{\phantom{a}bc}^{a}$ is the disformation
tensor, $^{4}\!\mathring{R}$ and $\mathring{\nabla}_{a}$ are the
Ricci scalar and covariant derivative adapted to the Levi-Civita connection,
respectively. (\ref{Lag4}) can be split into two parts
\begin{equation}
\mathcal{L}_{4}^{\text{H}}=\mathring{\mathcal{L}}_{4}^{\text{H}}+\tilde{\mathcal{L}}_{4}^{\text{H}},
\end{equation}
where 
\begin{equation}
\mathring{\mathcal{L}}_{4}^{\text{H}}=G_{4}{}^{4}\!\mathring{R}+G_{4,X}\left[\left(\mathring{\square}\phi\right)^{2}-\mathring{\nabla}^{a}\mathring{\nabla}^{b}\phi\mathring{\nabla}_{a}\mathring{\nabla}_{b}\phi\right],
\end{equation}
involves only the Levi-Civita connection, and terms involving the nonmetricity tensor are
\begin{eqnarray}
\tilde{\mathcal{L}}_{4}^{\text{H}} & \simeq & G_{4}\Big(L_{\phantom{a}ba}^{a}L_{\phantom{bc}c}^{bc}-L_{\phantom{c}bc}^{a}L_{\phantom{db}a}^{bc}\nonumber \\
 &  & \qquad-2\mathring{\square}\phi L_{\phantom{ca}a}^{ca}\nabla_{c}\phi+2\mathring{\nabla}^{a}\mathring{\nabla}^{b}\phi L_{\phantom{c}ba}^{c}\nabla_{c}\phi\nonumber \\
 &  & \qquad+L_{\phantom{ca}a}^{ca}\nabla_{c}\phi L_{\phantom{db}b}^{db}\nabla_{d}\phi-L^{dba}\nabla_{d}\phi L_{\phantom{c}ba}^{c}\nabla_{c}\phi\Big)\nonumber \\
 &  & -\left(G_{4,\phi}\mathring{\nabla}_{c}\phi-G_{4,X}\mathring{\nabla}_{c}\mathring{\nabla}_{d}\phi\mathring{\nabla}^{d}\phi\right)\left(L_{\phantom{cb}b}^{cb}-L_{\phantom{bc}b}^{bc}\right).
\end{eqnarray}

According to (\ref{cd1sca_dec}), in the unitary gauge, $\nabla_{a}\phi=-\frac{1}{N}u_{a}$
with $\frac{1}{N}\equiv\sqrt{2X}$. Thus $\pounds_{\bm{n}}\phi\rightarrow\pounds_{\bm{u}}\phi=\frac{1}{N}$.
The decomposition of the second derivative of the scalar field is
given in (\ref{cd2sca_dec}), from which we get
\begin{align}
\mathring{\nabla}_{a}\mathring{\nabla}_{b}\phi & =\frac{1}{N}\left(-n_{a}n_{b}\pounds_{\bm{u}}\ln N+2n_{(a}a_{b)}-K_{ab}\right)
\end{align}
in the unitary gauge. After some manipulations, we arrive at
\begin{equation}
\mathcal{L}_{4}^{\text{H(u.g.)}}=\mathring{\mathcal{L}}_{4}^{\text{H(u.g.)}}+\tilde{\mathcal{L}}_{4}^{\text{H(u.g.)}},\label{L4Hug}
\end{equation}
where \cite{Gleyzes:2013ooa,Fujita:2015ymn}
\begin{equation}
\mathring{\mathcal{L}}_{4}^{\text{H(u.g.)}}\simeq G_{4}\left(K^{ab}K_{ab}-K^{2}+\mathring{R}\right)-\frac{2}{N}G_{4,\phi}K+NG_{4,N}\left(K_{ab}K^{ab}-K^{2}\right),\label{L4Hug1}
\end{equation}
and
\begin{eqnarray}
\tilde{\mathcal{L}}_{4}^{\text{H(u.g.)}} & \simeq & G_{4}\Big[-L^{(2)}{}_{a}L^{(2)}{}^{a}-L^{(2)}{}^{a}L^{(3)}{}_{a}+L^{(4)}{}_{ab}L^{(4)}{}^{ba}\nonumber \\
 &  & \quad-L^{(4)}{}^{b}{}_{b}L^{(5)}{}_{\phantom{a}a}^{a}+2L^{(4)}{}^{ab}L^{(5)}{}_{ab}+L^{(1)}\left(L^{(4)}{}_{\phantom{a}a}^{a}+L^{(5)}{}_{\phantom{a}a}^{a}\right)\nonumber \\
 &  & \quad-L^{(2)}{}_{b}L^{(6)}{}_{\phantom{ba}a}^{ba}-L^{(3)}{}^{b}L^{(6)}{}_{\phantom{a}ba}^{a}+L^{(6)}{}_{\phantom{ca}a}^{ca}L^{(6)}{}_{\phantom{b}bc}^{b}-L^{(6)}{}_{\ \ }^{bac}L^{(6)}{}_{abc}\Big]\nonumber \\
 &  & +\frac{1}{N}G_{4,\phi}\left(L^{(5)}{}_{\phantom{a}a}^{a}-L^{(4)}{}_{\phantom{a}a}^{a}\right)\nonumber \\
 &  & +NG_{4,N}\Big[-\left(L^{(4)}{}_{\phantom{a}a}^{a}+L^{(5)}{}_{\phantom{a}a}^{a}\right)\pounds_{\bm{u}}\ln N-2KL^{(1)}+2\left(Kh^{ab}-K^{ab}\right)L_{ab}^{(5)}\nonumber \\
 &  & +a^{a}\left(3L_{a}^{(2)}+L_{a}^{(3)}-L^{(6)}{}_{ab}^{\phantom{ab}b}+L^{(6)}{}_{\phantom{b}ba}^{b}\right)\nonumber \\
 &  & -2L_{a}^{(2)}L^{(2)}{}^{a}+2L^{(1)}L^{(5)}{}_{\phantom{a}a}^{a}-\left(L^{(5)}{}_{\phantom{a}a}^{a}\right)^{2}+L^{(5)}{}^{ab}L_{ab}^{(5)}\Big].\label{L4Hug2}
\end{eqnarray}
In the above $G_{4}$ is understood as a function of $\phi$ and $N$.
Due to the presence of
\begin{equation}
NG_{4,N}\left(L^{(4)}{}_{\phantom{a}a}^{a}+L^{(5)}{}_{\phantom{a}a}^{a}\right)\pounds_{\bm{u}}\ln N
\end{equation}
in $\tilde{\mathcal{L}}_{4}^{\text{H(u.g.)}}$, after integrating
out the nonmetricity tensor (or equivalently, the disformation tensor),
the resulting Lagrangian involves $\left(\pounds_{\bm{u}}\ln N\right)^{2}$, signaling the existence of extra degrees of freedom. This is also consistent
with the analysis in \cite{Helpin:2019kcq,Kubota:2020ehu}.

\section{Spatially covariant monomials} \label{sec:mono}

According to the above discussion and especially the decomposition
of derivatives of the scalar field, the basic building blocks of the
nonmetricity theory coupled with a scalar field respecting the spatial
covariance are the usual quantities in the 3+1 decomposition, i.e.,
the lapse function $N$ and the spatial metric $h_{ab}$, the intrinsic
and extrinsic curvature $^{3}\!R_{ab}$ and $K_{ab}$ respectively,
as well as the projections of the disformation tensor $L^{(1)},\cdots,L_{cab}^{(6)}$
defined in (\ref{projdis}). Thus a general Lagrangian takes the form
\begin{align}
\mathcal{L} & =\mathcal{L}\left(\phi,N,{}^{3}\!R_{ab},h_{ab},K_{ab},L^{(1)},L_{a}^{(2)},L_{a}^{(3)},L_{ab}^{(4)},L_{ab}^{(5)},L_{cab}^{(6)};\mathrm{D}_{a},\pounds_{\bm{n}}\right),
\end{align}
where $L^{(i)},i=1,\cdots,5$ are introduced as a result of decomposition
of derivatives on the scalar fields in (\ref{cd2sca_dec}). According
to (\ref{DDphi_rel_xpl}), the spatial derivative $\mathrm{D}_{a}$
can also be split into $\mathring{\mathrm{D}}_{a}$ that is adapted
with the spatial metric $h_{ab}$ as well as contributions from the
nonmetricity tensor (or equivalently the disformation tensor). Similarly,
$^{3}\!R_{ab}$ can be further split into $^{3}\!\mathring{R}_{ab}$
and terms involving the disformation tensor (see Appendix \ref{app:deccurv}
for details). It is thus equivalent to consider the Lagrangian

\begin{align}
\mathcal{L} & =\mathcal{L}\left(\phi,N,{}^{3}\!\mathring{R}_{ab},h_{ab},K_{ab},L^{(1)},L_{a}^{(2)},L_{a}^{(3)},L_{ab}^{(4)},L_{ab}^{(5)},L_{cab}^{(6)};\mathring{\mathrm{D}}_{a},\pounds_{\bm{n}}\right),\label{LagSCG_gen}
\end{align}
which is more convenient when comparing with the usual SCG theory
without the nonmetricity tensor.

In the following, we will classify the spatially covariant scalar-nonmetricity
monomials. In particular, we will exhaust the monomials up to $d=3$
with $d$ the total number of derivatives in the unitary gauge. 

Since the number of monomials dramatically increases when $d$ goes
large. it is convenient to make a classification of these monomials.
To this end, we follow the approach developed in \cite{Gao:2020juc,Gao:2020yzr},
by classifying the SCG monomials according to their corresponding generally
covariant scalar-tensor (GST) expressions. The correspondence between
the SCG and GST monomials has been discussed in \cite{Gao:2020qxy,Gao:2020yzr,Hu:2021bbo}.
Schematically, we have
\begin{align}
K_{ij} & \sim a_{i}\sim L\sim\frac{1}{\nabla\phi}\nabla\nabla\phi,\quad{}^{3}\!R_{ij}\sim\frac{\left(\nabla\nabla\phi\right)^{2}}{\left(\nabla\phi\right)^{2}},\label{corr}
\end{align}
etc. Each GST monomial takes the general structure \cite{Gao:2020juc,Gao:2020yzr}
\begin{align}
\underset{c_{0}}{\underbrace{\cdots R\cdots}}\underset{c_{1}}{\underbrace{\cdots\nabla R\cdots}}\underset{c_{2}}{\underbrace{\cdots\nabla\nabla R\cdots}}\cdots\underset{d_{1}}{\underbrace{\cdots\nabla\phi\cdots}}\underset{d_{2}}{\underbrace{\cdots\nabla\nabla\phi\cdots}}\underset{d_{3}}{\underbrace{\cdots\nabla\nabla\nabla\phi\cdots}} & \cdots.
\end{align}
We label such a monomial with a set of integers $(c_{0},c_{1},c_{2},\cdots;d_{1},d_{2},d_{3},\cdots)$,
where $c_{i}$ is the number of the $i$-th order derivatives of the
curvature tensor and $d_{i}$ is the number of the $i$-th order derivatives
of the scalar field. Since the first-order derivative would not affect
the degeneracy structure of the theory, we suppress $d_{1}$ and use
the integer set $(c_{0},c_{1},c_{2},\cdots;d_{2},d_{3},d_{4},\cdots)$.
According to (\ref{corr}), in the GST correspondence each higher
derivative of the scalar field must be divided by $\nabla\phi$, we
define
\begin{align}
d & =\sum_{n=0}[(n+2)c_{n}+(n+1)d_{n+2}],
\end{align}
which is the total number of derivatives in the corresponding SCG
monomials.

The appearance of the Lie derivative in the Lagrangian could alter
the dynamics of the theory and possibly cause non-physical ghosts.
For example, as we have discussed above, the first-order Lie derivatives
of the disformation tensor will result in the higher derivatives of
the scalar field (see (\ref{cd3sca_dec})). For this reason, in this
work we do not consider the Lie derivatives explicitly in the Lagrangian
in order to avoid ghostlike degrees of freedom. In other words, the
Lie derivative only enters implicitly in the extrinsic curvature $\pounds_{\bm{n}}h_{ab}=2K_{ab}$.
Hence the Lagrangian takes the general form\footnote{In principle, in spatially covariant gravity, the lapse function, spatial metric as well as the nonmetricity tensor should be treated on equal footing. Therefore, it is natural to investigate how
to build the theory with Lie derivatives such as $\pounds_{\bm{n}}N$,
$\pounds_{\bm{n}}L$ etc. The case without the nonmetricity tensor
has been discussed in \cite{Gao:2018znj}.}
\begin{align}
\mathcal{L} & =\mathcal{L}(\phi,N,{}^{3}\!\mathring{R}_{ij},h_{ij},K_{ij},L^{(1)},L_{i}^{(2)},L_{i}^{(3)},L_{ij}^{(4)},L_{ij}^{(5)},L_{kij}^{(6)};\mathring{\mathrm{D}}_{i}).\label{eq:3.10}
\end{align}
In Table \ref{tab:dim}, we list all the building blocks in their
schematic form of SCG with nonmetricity. 

\begin{table}[H]
\begin{centering}
\begin{tabular}{|c|c|c|c|c|c|c|c|}
\hline 
{\footnotesize{}$d$} & {\footnotesize{}$\#_{\text{D}}$} & {\footnotesize{}Form} & {\footnotesize{}$(c_{0},c_{1},c_{2};d_{2},d_{3},d_{4})$} & {\footnotesize{}$d$} & {\footnotesize{}$\#_{\text{D}}$} & {\footnotesize{}Form} & {\footnotesize{}$(c_{0},c_{1},c_{2};d_{2},d_{3},d_{4})$}\tabularnewline
\hline 
\multirow{3}{*}{{\footnotesize{}1}} & \multirow{3}{*}{{\footnotesize{}0}} & {\footnotesize{}$K$} & \multirow{3}{*}{{\footnotesize{}$(0,0,0;1,0,0)$}} & \multirow{4}{*}{{\footnotesize{}3}} & {\footnotesize{}1} & {\footnotesize{}$\mathring{\mathrm{D}}\,{}^{3}\!\mathring{R}$} & {\footnotesize{}$(0,1,0;0,0,0)$}\tabularnewline
\cline{3-3} \cline{6-8} \cline{7-8} \cline{8-8} 
 &  & {\footnotesize{}$a$} &  &  & \multirow{3}{*}{{\footnotesize{}2}} & {\footnotesize{}$\mathring{\mathrm{D}}\mathring{\mathrm{D}}K$} & \multirow{3}{*}{{\footnotesize{}$(0,0,0;0,0,1)$}}\tabularnewline
\cline{3-3} \cline{7-7} 
 &  & {\footnotesize{}$L$} &  &  &  & {\footnotesize{}$\mathring{\mathrm{D}}\mathring{\mathrm{D}}a$} & \tabularnewline
\cline{1-4} \cline{2-4} \cline{3-4} \cline{4-4} \cline{7-7} 
\multirow{4}{*}{{\footnotesize{}2}} & {\footnotesize{}0} & {\footnotesize{}$^{3}\!\mathring{R}$} & {\footnotesize{}$(1,0,0;0,0,0)$} &  &  & {\footnotesize{}$\mathring{\mathrm{D}}\mathring{\mathrm{D}}L$} & \tabularnewline
\cline{2-8} \cline{3-8} \cline{4-8} \cline{5-8} \cline{6-8} \cline{7-8} \cline{8-8} 
 & \multirow{3}{*}{{\footnotesize{}1}} & {\footnotesize{}$\mathring{\mathrm{D}}K$} & \multirow{3}{*}{{\footnotesize{}$(0,0,0;0,1,0)$}} & \multirow{4}{*}{{\footnotesize{}4}} & {\footnotesize{}2} & {\footnotesize{}$\mathring{\mathrm{D}}\mathring{\mathrm{D}}\,{}^{3}\!\mathring{R}$} & {\footnotesize{}$(0,0,1;0,0,0)$}\tabularnewline
\cline{3-3} \cline{6-8} \cline{7-8} \cline{8-8} 
 &  & {\footnotesize{}$\mathring{\mathrm{D}}a$} &  &  & \multirow{3}{*}{{\footnotesize{}3}} & {\footnotesize{}$\mathring{\mathrm{D}}\mathring{\mathrm{D}}\mathring{\mathrm{D}}K$} & \multirow{3}{*}{{\footnotesize{}5th der.}}\tabularnewline
\cline{3-3} \cline{7-7} 
 &  & {\footnotesize{}$\mathring{\mathrm{D}}L$} &  &  &  & {\footnotesize{}$\mathring{\mathrm{D}}\mathring{\mathrm{D}}\mathring{\mathrm{D}}a$} & \tabularnewline
\cline{1-4} \cline{2-4} \cline{3-4} \cline{4-4} \cline{7-7} 
\multirow{1}{*}{} &  &  &  &  &  & {\footnotesize{}$\mathring{\mathrm{D}}\mathring{\mathrm{D}}\mathring{\mathrm{D}}L$} & \tabularnewline
\hline 
\end{tabular}
\par\end{centering}
\caption{Classification of basic building blocks up to $d=4$.\label{tab:dim}}
\end{table}

In the following, we provide the explicit expressions for all the scalar
monomials up to $d=3$. We note that only $c_{0}$, $d_{2}$ and $d_{3}$
are needed for our purpose and thus suppress $c_{1}$, $c_{2}$ and
$d_{4}$ for simplicity. For $d=1$, there are 4 monomials, which
are listed in Table \ref{tab:d1}.

\begin{table}[H]
	\begin{centering}
		\begin{tabular}{|c|c|c|}
			\hline 
			Form & Explicit form & {\footnotesize{}$(c_{0};d_{2},d_{3})$}\tabularnewline
			\hline 
			{\footnotesize{}$K$} & {\footnotesize{}$K$} & \multirow{2}{*}{{\footnotesize{}$(0;1,0)$}}\tabularnewline
			\cline{1-2} \cline{2-2} 
			{\footnotesize{}$L$} & {\footnotesize{}$L^{(1)},L_{\ \ \ i}^{(4)\ i},L_{\ \ \ i}^{(5)\ i}$} & \tabularnewline
			\hline 
		\end{tabular}
		\par\end{centering}
	\caption{Monomials with $d=1$.\label{tab:d1}}
\end{table}

\begin{table}[H]
\begin{centering}
{\footnotesize{}}%
\begin{tabular}{|c|c|c|}
\hline 
{\scriptsize{}Form} & {\scriptsize{}Monomials} & {\scriptsize{}$(c_{0};d_{2},d_{3})$}\tabularnewline
\hline 
{\scriptsize{}$^{3}\!\overset{\circ}{R}$} & {\scriptsize{}$^{3}\!\mathring{R}$} & {\scriptsize{}$(1;0,0)$}\tabularnewline
\hline 
{\scriptsize{}$K^{2}$} & {\scriptsize{}$K^{ij}K_{ij}$} & \multirow{7}{*}{{\scriptsize{}$(0;2,0)$}}\tabularnewline
\cline{1-2} \cline{2-2} 
\multirow{1}{*}{{\scriptsize{}$KL$}} & {\scriptsize{}$K^{ij}L_{ij}^{(4)},K^{ij}L_{ij}^{(5)}$} & \tabularnewline
\cline{1-2} \cline{2-2} 
{\scriptsize{}$aa$} & {\scriptsize{}$a_{i}a^{i}$} & \tabularnewline
\cline{1-2} \cline{2-2} 
\multirow{1}{*}{{\scriptsize{}$aL$}} & {\scriptsize{}$a^{i}L_{i}^{(2)},a^{i}L_{i}^{(3)},a^{i}L_{\ \ \ i\ j}^{(6)\ j},a^{i}L_{\ \ \ \ ij}^{(6)j}$} & \tabularnewline
\cline{1-2} \cline{2-2} 
\multirow{3}{*}{{\scriptsize{}$LL$}} & {\scriptsize{}$L_{i}^{(2)}L^{(2)i},L_{i}^{(2)}L^{(3)i},L_{i}^{(2)}L_{\ \ \ \ \ j}^{(6)ij},L_{i}^{(2)}L_{\ \ \ \ \ j}^{(6)ji},L_{i}^{(3)}L^{(3)i},L_{i}^{(3)}L_{\ \ \ \ \ j}^{(6)ij},$} & \tabularnewline
 & {\scriptsize{}$L_{i}^{(3)}L_{\ \ \ \ \ j}^{(6)ji},L_{\ \ \ i\ k}^{(6)\ k}L_{\ \ \ \ \ j}^{(6)ij},L_{\ \ \ i\ k}^{(6)\ k}L_{\ \ \ \ \ j}^{(6)ji},L_{\ \ \ \ ik}^{(6)k}L_{\ \ \ \ \ j}^{(6)ji},$} & \tabularnewline
 & {\scriptsize{}$L_{ij}^{(4)}L^{(4)ij},L_{ij}^{(4)}L^{(4)ji},L_{ij}^{(4)}L^{(5)ij},L_{ij}^{(5)}L^{(5)ij},L_{kij}^{(6)}L^{(6)kij},L_{kij}^{(6)}L^{(6)ikj}$} & \tabularnewline
\hline 
\end{tabular}{\footnotesize\par}
\par\end{centering}

\caption{Unfactorizable and irreducible monomials with $d=2$.\label{tab:d2}}
\end{table}

For $d=2$, the number of monomials dramatically increases. For the
sake of brevity, we follow the formalism developed in \cite{Gao:2020qxy}
and concentrate on the unfactorizable and irreducible monomials, which
are listed in Table \ref{tab:d2}. That is, these monomials are not products
of more than one scalar monomial and cannot be reduced by integrations
by parts, i.e., not total derivatives nor linear combinations of other
irreducible ones. Terms in the form $\mathring{\mathrm{D}}a$ and
$\mathring{\mathrm{D}}L$ are clearly total derivatives. There are
two special categories with $\left(c_{0};d_{2},d_{3}\right)=\left(0;0,1\right)$
in the form $\mathring{\mathrm{D}}a$ and $\mathring{\mathrm{D}}L$,
which we do not list in Table \ref{tab:d2} and are total derivatives
at order $d=2$, but will contribute to the factorizable monomials
with $d=3$\footnote{This is similar to the case in \cite{Gao:2020qxy} (see eq. (2.56)).}.
There is only one monomial in the form $\mathring{\mathrm{D}}a$,
\begin{equation}
\mathring{\mathrm{D}}^{i}a_{i},
\end{equation}
and 4 monomials in the form $\mathring{\mathrm{D}}L$,
\begin{equation}
\ensuremath{\mathring{\mathrm{D}}^{i}L_{i}^{(2)},\mathring{\mathrm{D}}^{i}L_{i}^{(3)},\mathring{\mathrm{D}}^{i}L_{\ \ \ i\ j}^{(6)\ j},\mathring{\mathrm{D}}^{i}L_{\ \ \ \ ij}^{(6)j}}.
\end{equation}

The unfactorizable and irreducible monomials with $d=3$ are listed in Table \ref{tab:d3}. Terms in the form $\mathring{\mathrm{D}}\mathring{\mathrm{D}}K$,
$\mathring{\mathrm{D}}\mathring{\mathrm{D}}L$, $\mathring{\mathrm{D}}aK$,
$\mathring{\mathrm{D}}LK$, $\mathring{\mathrm{D}}La$ are clearly
reducible.

{\footnotesize{}}
\begin{table}[H]
\begin{centering}
{\footnotesize{}}%
\begin{tabular}{|c|c|c|}
\hline 
{\scriptsize{}Form} & {\scriptsize{}Monomials} & {\scriptsize{}$(c_{0};d_{2},d_{3})$}\tabularnewline
\hline 
{\scriptsize{}$\overset{\circ}{R}K$} & {\scriptsize{}$^{3}\!\mathring{R}_{ij}K^{ij}$} & \multirow{2}{*}{{\scriptsize{}$(1;1,0)$}}\tabularnewline
\cline{1-2} \cline{2-2} 
{\scriptsize{}$\overset{\circ}{R}L$} & {\scriptsize{}$^{3}\!\mathring{R}^{ij}L_{ij}^{(4)},{}^{3}\!\mathring{R}^{ij}L_{ij}^{(5)}$} & \tabularnewline
\hline 
{\scriptsize{}$K\mathring{\mathrm{D}}a$} & {\scriptsize{}$K^{ij}\mathring{\mathrm{D}}_{i}a_{j}$} & \multirow{7}{*}{{\scriptsize{}$(0;1,1)$}}\tabularnewline
\cline{1-2} \cline{2-2} 
\multirow{1}{*}{{\scriptsize{}$\left(\mathring{\mathrm{D}}K\right)L$}} & {\scriptsize{}$\mathring{\mathrm{D}}^{i}K_{i}^{\ j}L_{j}^{(2)},\mathring{\mathrm{D}}^{i}K_{i}^{\ j}L_{j}^{(3)},\mathring{\mathrm{D}}_{k}K_{ij}L^{(6)kij},\mathring{\mathrm{D}}_{k}K_{ij}L^{(6)ikj},\mathring{\mathrm{D}}^{i}K_{i}^{\ j}L^{(6)}{}_{\ jk}^{k},\mathring{\mathrm{D}}^{i}K_{i}^{\ j}L^{(6)}{}_{j\ k}^{\ k}$} & \tabularnewline
\cline{1-2} \cline{2-2} 
{\scriptsize{}$\left(\mathring{\mathrm{D}}a\right)L$} & {\scriptsize{}$\mathring{\mathrm{D}}_{i}a_{j}L^{(4)ij},\mathring{\mathrm{D}}_{i}a_{j}L^{(5)ij}$} & \tabularnewline
\cline{1-2} \cline{2-2} 
\multirow{4}{*}{{\scriptsize{}$\left(\mathring{\mathrm{D}}L\right)L$}} & {\scriptsize{}$\mathring{\mathrm{D}}^{i}L^{(2)j}L_{ij}^{(4)},\mathring{\mathrm{D}}^{i}L^{(2)j}L_{ji}^{(4)},\mathring{\mathrm{D}}^{i}L^{(2)j}L_{ij}^{(5)},\mathring{\mathrm{D}}^{i}L^{(3)j}L_{ij}^{(4)},\mathring{\mathrm{D}}^{i}L^{(3)j}L_{ji}^{(4)},$} & \tabularnewline
 & {\scriptsize{}$\mathring{\mathrm{D}}^{i}L^{(3)j}L_{ij}^{(5)},\mathring{\mathrm{D}}^{i}L_{\ \ \ ij}^{(4)}L_{\ \ \ k}^{(6)\ jk},\mathring{\mathrm{D}}^{i}L_{\ \ \ ij}^{(4)}L_{\ \ \ \ \ k}^{(6)jk},\mathring{\mathrm{D}}^{i}L_{\ \ \ ji}^{(4)}L_{\ \ \ k}^{(6)\ jk},\mathring{\mathrm{D}}^{i}L_{\ \ \ ji}^{(4)}L_{\ \ \ \ \ k}^{(6)jk},$} & \tabularnewline
 & {\scriptsize{}$\mathring{\mathrm{D}}_{k}L_{ij}^{(4)}L^{(6)kij},\mathring{\mathrm{D}}_{k}L_{ij}^{(4)}L^{(6)ikj},\mathring{\mathrm{D}}_{k}L_{ij}^{(4)}L^{(6)jki},\mathring{\mathrm{D}}^{i}L_{\ \ \ ij}^{(5)}L_{\ \ \ k}^{(6)\ jk},$} & \tabularnewline
 & {\scriptsize{}$\mathring{\mathrm{D}}^{i}L_{\ \ \ ij}^{(5)}L_{\ \ \ \ \ k}^{(6)jk},\mathring{\mathrm{D}}_{k}L_{ij}^{(5)}L^{(6)kij},\mathring{\mathrm{D}}_{k}L_{ij}^{(5)}L^{(6)ikj}$} & \tabularnewline
\hline 
{\scriptsize{}$K^{3}$} & {\scriptsize{}$K^{ij}K_{jk}K_{i}^{k}$} & \multirow{31}{*}{{\scriptsize{}$(0;3,0)$}}\tabularnewline
\cline{1-2} \cline{2-2} 
{\scriptsize{}$K^{2}L$} & {\scriptsize{}$K^{ik}K_{k}^{j}L_{ij}^{(4)},K^{ik}K_{k}^{j}L_{ij}^{(5)}$} & \tabularnewline
\cline{1-2} \cline{2-2} 
{\scriptsize{}$Ka^{2}$} & {\scriptsize{}$K^{ij}a_{i}a_{j}$} & \tabularnewline
\cline{1-2} \cline{2-2} 
\multirow{8}{*}{{\scriptsize{}$KL^{2}$}} & {\scriptsize{}$K^{ij}L_{i}^{(2)}L_{j}^{(2)},K^{ij}L_{i}^{(2)}L_{j}^{(3)},K^{ij}L_{i}^{(2)}L_{\ \ \ j\ k}^{(6)\ k},K^{ij}L_{i}^{(2)}L_{\ \ \ \ jk}^{(6)k},K^{ij}L_{i}^{(3)}L_{j}^{(3)},$} & \tabularnewline
 & {\scriptsize{}$K^{ij}L_{i}^{(3)}L_{\ \ \ j\ k}^{(6)\ k},K^{ij}L_{i}^{(3)}L_{\ \ \ \ jk}^{(6)k},K^{ij}L_{\ \ \ i\ k}^{(6)\ k}L_{\ \ \ j\ l}^{(6)\ l},K^{ij}L_{\ \ \ i\ k}^{(6)\ k}L_{\ \ \ \ jl}^{(6)l},$} & \tabularnewline
 & {\scriptsize{}$K^{ij}L_{\ \ \ \ ik}^{(6)k}L_{\ \ \ \ jl}^{(6)l}$} & \tabularnewline
\cline{2-2} 
 & {\scriptsize{}$K^{ij}L^{(2)k}L_{\ \ \ ijk}^{(6)},K^{ij}L^{(2)k}L_{\ \ \ kij}^{(6)},K^{ij}L^{(3)k}L_{\ \ \ ijk}^{(6)},K^{ij}L^{(3)k}L_{\ \ \ kij}^{(6)},$} & \tabularnewline
 & {\scriptsize{}$K^{ij}L_{\ \ \ \ \ l}^{(6)kl}L_{\ \ \ ijk}^{(6)},K^{ij}L_{\ \ \ \ \ l}^{(6)kl}L_{\ \ \ kij}^{(6)},K^{ij}L_{\ \ \ \ \ l}^{(6)lk}L_{\ \ \ ijk}^{(6)},K^{ij}L_{\ \ \ \ \ l}^{(6)lk}L_{\ \ \ kij}^{(6)}$} & \tabularnewline
\cline{2-2} 
 & {\scriptsize{}$K^{ij}L_{\ \ \ i}^{(4)\ k}L_{\ \ \ jk}^{(4)},K^{ij}L_{\ \ \ i}^{(4)\ k}L_{\ \ \ kj}^{(4)},K^{ij}L_{\ \ \ \ i}^{(4)k}L_{\ \ \ kj}^{(4)}$} & \tabularnewline
 & {\scriptsize{}$K^{ij}L_{\ \ \ i}^{(4)\ k}L_{\ \ \ kj}^{(5)},K^{ij}L_{\ \ \ \ i}^{(4)k}L_{\ \ \ kj}^{(5)},K^{ij}L_{\ \ \ i}^{(5)\ k}L_{\ \ \ jk}^{(5)}$} & \tabularnewline
\cline{2-2} 
 & {\scriptsize{}$K^{ij}L_{\ \ \ i}^{(6)\ kl}L_{\ \ \ jkl}^{(6)},K^{ij}L_{\ \ \ i}^{(6)\ kl}L_{\ \ \ kjl}^{(6)},K^{ij}L_{\ \ \ \ i}^{(6)k\ l}L_{\ \ \ kjl}^{(6)},K^{ij}L_{\ \ \ \ i}^{(6)l\ k}L_{\ \ \ kjl}^{(6)}$} & \tabularnewline
\cline{1-2} \cline{2-2} 
\multirow{1}{*}{{\scriptsize{}$KaL$}} & {\scriptsize{}$K^{ij}a_{k}L_{\ \ \ \ ij}^{(6)k},K^{ij}a_{k}L_{\ \ \ i\ j}^{(6)\ k},K^{ij}a_{i}L_{j}^{(2)},K^{ij}a_{i}L_{j}^{(3)},K^{ij}a_{i}L_{\ \ \ j\ k}^{(6)\ k},K^{ij}a_{i}L_{\ \ \ \ jk}^{(6)k}$} & \tabularnewline
\cline{1-2} \cline{2-2} 
{\scriptsize{}$a^{2}L$} & {\scriptsize{}$a^{i}a^{j}L_{ij}^{(4)},a^{i}a^{j}L_{ij}^{(5)}$} & \tabularnewline
\cline{1-2} \cline{2-2} 
\multirow{4}{*}{{\scriptsize{}$aL^{2}$}} & {\scriptsize{}$a^{i}L_{ik}^{(4)}L^{(2)k},a^{i}L_{ik}^{(4)}L^{(3)k},a^{i}L_{ik}^{(4)}L_{\ \ \ \ \ j}^{(6)kj},a^{i}L_{ik}^{(4)}L_{\ \ \ \ \ j}^{(6)jk},$} & \tabularnewline
 & {\scriptsize{}$a^{i}L_{ki}^{(4)}L^{(2)k},a^{i}L_{ki}^{(4)}L^{(3)k},a^{i}L_{ki}^{(4)}L_{\ \ \ \ \ j}^{(6)kj},a^{i}L_{ki}^{(4)}L_{\ \ \ \ \ j}^{(6)jk},a^{i}L_{ik}^{(5)}L^{(2)k},a^{i}L_{ik}^{(5)}L^{(3)k},$} & \tabularnewline
 & {\scriptsize{}$a^{i}L_{ik}^{(5)}L_{\ \ \ \ \ j}^{(6)kj},a^{i}L_{ik}^{(5)}L_{\ \ \ \ \ j}^{(6)jk},a^{i}L^{(4)kl}L_{\ \ \ ikl}^{(6)},a^{i}L^{(4)kl}L_{\ \ \ kil}^{(6)},a^{i}L^{(4)kl}L_{\ \ \ lik}^{(6)},$} & \tabularnewline
 & {\scriptsize{}$a^{i}L^{(5)kl}L_{\ \ \ ikl}^{(6)},a^{i}L^{(5)kl}L_{\ \ \ kil}^{(6)}$} & \tabularnewline
\cline{1-2} \cline{2-2} 
\multirow{14}{*}{{\scriptsize{}$L^{3}$}} & {\scriptsize{}$L^{(2)i}L^{(2)j}L_{ij}^{(4)},L^{(2)i}L^{(3)j}L_{ij}^{(4)},L^{(2)i}L_{\ \ \ \ \ k}^{(6)jk}L_{ij}^{(4)},L^{(2)i}L_{\ \ \ \ \ k}^{(6)kj}L_{ij}^{(4)},L^{(2)i}L^{(3)j}L_{ji}^{(4)},$} & \tabularnewline
 & {\scriptsize{}$L^{(2)i}L_{\ \ \ \ \ k}^{(6)jk}L_{ji}^{(4)},L^{(2)i}L_{\ \ \ \ \ k}^{(6)kj}L_{ji}^{(4)},L^{(3)i}L^{(3)j}L_{ij}^{(4)},L^{(3)i}L_{\ \ \ \ \ k}^{(6)jk}L_{ij}^{(4)},L^{(3)i}L_{\ \ \ \ \ k}^{(6)kj}L_{ij}^{(4)},$} & \tabularnewline
 & {\scriptsize{}$L^{(3)i}L_{\ \ \ \ \ k}^{(6)jk}L_{ji}^{(4)},L^{(3)i}L_{\ \ \ \ \ k}^{(6)kj}L_{ji}^{(4)},L_{\ \ \ \ \ k}^{(6)ik}L_{\ \ \ \ \ l}^{(6)jl}L_{ij}^{(4)},L_{\ \ \ \ \ k}^{(6)ik}L_{\ \ \ \ \ l}^{(6)lj}L_{ij}^{(4)},$} & \tabularnewline
 & {\scriptsize{}$L_{\ \ \ \ \ k}^{(6)ik}L_{\ \ \ \ \ l}^{(6)lj}L_{ji}^{(4)},L_{\ \ \ \ \ k}^{(6)ki}L_{\ \ \ \ \ l}^{(6)lj}L_{ij}^{(4)},L^{(2)i}L^{(2)j}L_{ij}^{(5)},L^{(2)i}L^{(3)j}L_{ij}^{(5)},L^{(2)i}L_{\ \ \ \ \ k}^{(6)jk}L_{ij}^{(5)},$} & \tabularnewline
 & {\scriptsize{}$L^{(2)i}L_{\ \ \ \ \ k}^{(6)kj}L_{ij}^{(5)},L^{(3)i}L^{(3)j}L_{ij}^{(5)},L^{(3)i}L_{\ \ \ \ \ k}^{(6)jk}L_{ij}^{(5)},L^{(3)i}L_{\ \ \ \ \ k}^{(6)kj}L_{ij}^{(5)},L_{\ \ \ \ \ k}^{(6)ik}L_{\ \ \ \ \ l}^{(6)jl}L_{ij}^{(5)},$} & \tabularnewline
 & {\scriptsize{}$L_{\ \ \ \ \ k}^{(6)ik}L_{\ \ \ \ \ l}^{(6)lj}L_{ij}^{(5)},L_{\ \ \ \ \ k}^{(6)ki}L_{\ \ \ \ \ l}^{(6)lj}L_{ij}^{(5)},L_{k}^{(2)}L_{ij}^{(4)}L^{(6)kij},L_{k}^{(2)}L_{ij}^{(4)}L^{(6)ikj},L_{k}^{(2)}L_{ji}^{(4)}L^{(6)ikj},$} & \tabularnewline
 & {\scriptsize{}$L_{k}^{(2)}L_{ij}^{(5)}L^{(6)kij},L_{k}^{(2)}L_{ij}^{(5)}L^{(6)ikj},L_{k}^{(3)}L_{ij}^{(4)}L^{(6)kij},L_{k}^{(3)}L_{ij}^{(4)}L^{(6)ikj},L_{k}^{(3)}L_{ji}^{(4)}L^{(6)ikj},$} & \tabularnewline
 & {\scriptsize{}$L_{k}^{(3)}L_{ij}^{(5)}L^{(6)kij},L_{k}^{(3)}L_{ij}^{(5)}L^{(6)ikj},L_{\ \ \ k\ l}^{(6)\ l}L_{ij}^{(4)}L^{(6)kij},L_{\ \ \ k\ l}^{(6)\ l}L_{ij}^{(4)}L^{(6)ikj},$} & \tabularnewline
 & {\scriptsize{}$L_{\ \ \ k\ l}^{(6)\ l}L_{ij}^{(5)}L^{(6)kij},L_{\ \ \ k\ l}^{(6)\ l}L_{ij}^{(5)}L^{(6)ikj},L_{\ \ \ \ kl}^{(6)l}L_{ij}^{(4)}L^{(6)kij},L_{\ \ \ \ kl}^{(6)l}L_{ij}^{(4)}L^{(6)ikj},$} & \tabularnewline
 & {\scriptsize{}$L_{\ \ \ k\ l}^{(6)\ l}L_{ji}^{(4)}L^{(6)ikj},L_{\ \ \ \ kl}^{(6)l}L_{ji}^{(4)}L^{(6)ikj},L_{\ \ \ \ kl}^{(6)l}L_{ij}^{(5)}L^{(6)kij},L_{\ \ \ \ kl}^{(6)l}L_{ij}^{(5)}L^{(6)ikj},$} & \tabularnewline
 & {\scriptsize{}$L^{(4)ij}L_{\ \ \ kli}^{(6)}L_{\ \ \ \ \ j}^{(6)kl},L^{(4)ij}L_{\ \ \ ikl}^{(6)}L_{\ \ \ \ \ j}^{(6)kl},L^{(4)ij}L_{\ \ \ ikl}^{(6)}L_{\ \ \ j\ \ }^{(6)\ kl},L^{(4)ij}L_{\ \ \ kli}^{(6)}L_{\ \ \ \ \ j}^{(6)lk},$} & \tabularnewline
 & {\scriptsize{}$L^{(4)ji}L_{\ \ \ ikl}^{(6)}L_{\ \ \ \ \ j}^{(6)kl},L^{(5)ij}L_{\ \ \ kli}^{(6)}L_{\ \ \ \ \ j}^{(6)kl},L^{(5)ij}L_{\ \ \ ikl}^{(6)}L_{\ \ \ \ \ j}^{(6)kl},L^{(5)ij}L_{\ \ \ ikl}^{(6)}L_{\ \ \ j\ \ }^{(6)\ kl},$} & \tabularnewline
 & {\scriptsize{}$L^{(5)ij}L_{\ \ \ kli}^{(6)}L_{\ \ \ \ \ j}^{(6)lk},L_{ij}^{(4)}L^{(4)jk}L_{\ \ \ k}^{(4)\ i},L_{ij}^{(4)}L^{(4)jk}L_{\ \ \ \ k}^{(4)i},L_{ij}^{(4)}L^{(4)jk}L_{\ \ \ k}^{(5)\ i},$} & \tabularnewline
 & {\scriptsize{}$L_{ij}^{(4)}L^{(4)kj}L_{\ \ \ k}^{(5)\ i},L_{ij}^{(4)}L^{(5)jk}L_{\ \ \ k}^{(4)\ i},L_{ij}^{(4)}L^{(5)ik}L_{\ \ \ j}^{(5)\ k},L_{ij}^{(5)}L^{(5)ik}L_{\ \ \ \ k}^{(5)j}$} & \tabularnewline
\hline 
\end{tabular}{\footnotesize\par}
\par\end{centering}
{\footnotesize{}\caption{Unfactorizable and irreducible monomials with $d=3$. \label{tab:d3}}
}{\footnotesize\par}
\end{table}
{\footnotesize\par}

In this work, since we do not consider the Lie derivatives of the
disformation tensor, the disformation tensor enters the Lagrangian
as an auxiliary variable. After solving the disformation tensor in
terms of the metric variables (i.e., the lapse function, extrinsic
curvature and their spatial derivatives), we are left with an effective
SCG Lagrangian with modified coefficients. For $d=2$, since the disformation
tensor arises without any derivatives, the resulting effective SCG
Lagrangian can be obtained straightforwardly, which will be shown explicitly
in Sec. \ref{sec:quad}. For $d\geq3$, the appearance of spatial
derivatives of the disformation tensor makes solving the disformation
tensor involved. In particular, reversing the spatial derivatives
would introduce nonlocal-type operators, which are not included in
the original polynomial-type SCG Lagrangian.

In principle, the Lie derivatives of the disformation tensor (or some of its components) can be taken into account as the basic building blocks of the theory. In this case the disformation tensor is no longer an auxiliary field and the number of degrees of freedom will drastically increases. Nevertheless, this case will be of interest since in the framework of metric-affine gravity, one can treat the metric and the affine connection on an equal footing and thus introduce a kinetic term for the connection (e.g., the disformation tensor).

\section{The quadratic theory} \label{sec:quad}

In this section, we focus on the theory with $d=2$ as a concrete example, where the Lagrangian is quadratic in the extrinsic curvature,
the acceleration and the disformation tensor. The general Lagrangian
is a linear combination of all the monomials with $d=2$:
\begin{eqnarray}
\mathcal{L} & = & ^{3}\!\mathring{R}+K^{ij}K_{ij}-K^{2}+c_{1}KL^{(1)}+c_{2}KL_{\ \ \ i}^{(4)\ i}+c_{3}KL_{\ \ \ i}^{(5)\ i}+c_{4}K^{ij}L_{ij}^{(4)}+c_{5}K^{ij}L_{ij}^{(5)}\nonumber \\
 &  & +d_{1}L^{(1)}L^{(1)}+d_{2}L^{(1)}L_{\ \ \ i}^{(4)\ i}+d_{3}L^{(1)}L_{\ \ \ i}^{(5)\ i}+d_{4}L_{\ \ \ i}^{(4)\ i}L_{\ \ \ j}^{(4)\ j}+d_{5}L_{\ \ \ i}^{(4)\ i}L_{\ \ \ j}^{(5)\ j}\nonumber \\
 &  & +d_{6}L_{\ \ \ i}^{(5)\ i}L_{\ \ \ j}^{(5)\ j}+d_{7}L_{i}^{(2)}L^{(2)i}+d_{8}L_{i}^{(2)}L^{(3)i}+d_{9}L_{i}^{(2)}L_{\ \ \ \ \ j}^{(6)ij}+d_{10}L_{i}^{(2)}L_{\ \ \ \ \ j}^{(6)ji}\nonumber \\
 &  & +d_{11}L_{i}^{(3)}L^{(3)i}+d_{12}L_{i}^{(3)}L_{\ \ \ \ \ j}^{(6)ij}+d_{13}L_{i}^{(3)}L_{\ \ \ \ \ j}^{(6)ji}+d_{14}L_{\ \ \ i\ k}^{(6)\ k}L_{\ \ \ \ \ j}^{(6)ij}+d_{15}L_{\ \ \ i\ k}^{(6)\ k}L_{\ \ \ \ \ j}^{(6)ji}\nonumber \\
 &  & +d_{16}L_{\ \ \ \ ik}^{(6)k}L_{\ \ \ \ \ j}^{(6)ji}+d_{17}L_{ij}^{(4)}L^{(4)ij}+d_{18}L_{ij}^{(4)}L^{(4)ji}+d_{19}L_{ij}^{(4)}L^{(5)ij}+d_{20}L_{ij}^{(5)}L^{(5)ij}\nonumber \\
 &  & +d_{21}L_{kij}^{(6)}L^{(6)kij}+d_{22}L_{kij}^{(6)}L^{(6)ikj}+f_{1}a^{i}L_{i}^{(2)}+f_{2}a^{i}L_{i}^{(3)}+f_{3}a^{i}L_{\ \ \ i\ j}^{(6)\ j}+f_{4}a^{i}L_{\ \ \ \ ij}^{(6)j},\label{Lagd2}
\end{eqnarray}
where $c_{i},i=1,...,5$, $d_{i},i=1,...,22$ and $f_{i},i=1,...,4$
are arbitrary functions of $t$ and $N$. The first three terms in
(\ref{Lagd2}) correspond to the 3+1 decomposition of general relativity,
while the remaining terms are contributions from the disformation tensor.

As mentioned before, the disformation tensor does not contain
any time derivatives, and thus $L^{(i)},i=1,...,6$ are auxiliary variables
and can be solved by their equations of motion. Since the Lagrangian is quadratic
in the disformation tensor, the extrinsic curvature and the acceleration,
the solutions for $L^{(i)}$ must be linear in the extrinsic curvature
and the acceleration. We make the ansatz for the solutions as follows
\begin{align}
L^{(1)} & =m^{(1)}K,\label{eq:4.2}\\
L_{i}^{(2)} & =m^{(2)}a_{i},\label{eq:4.3}\\
L_{i}^{(3)} & =m^{(3)}a_{i},\label{eq:4.4}\\
L_{ij}^{(4)} & =m_{1}^{(4)}K_{ij}+m_{2}^{(4)}Kh_{ij},\label{eq:4.5}\\
L_{ij}^{(5)} & =m_{1}^{(5)}K_{ij}+m_{2}^{(5)}Kh_{ij},\label{eq:4.6}\\
L_{kij}^{(6)} & =m_{1}^{(6)}a_{k}h_{ij}+m_{2}^{(6)}a_{(i}h_{j)k}, \label{eq:4.7}
\end{align}
where $m^{(1)},m^{(2)},m^{(3)},m_{1}^{(4)},m_{2}^{(4)},m_{1}^{(5)},m_{2}^{(5)},m_{1}^{(6)}$
and $m_{2}^{(6)}$ are coefficients to be determined. The detailed
solutions for $m$'s can be found in Appendix \ref{app:soldisten},
which are functions of $c_{i}$, $d_{i}$ and $f_{i}$.

After plugging the solutions of $m$'s into the Lagrangian (\ref{Lagd2}),
we arrive at an effective SCG Lagrangian for $d=2$:
\begin{align}
\mathcal{L} & =\,{}^{3}\!\mathring{R}+pK^{ij}K_{ij}+qK^{2}+ra_{i}a^{i},\label{Lageff}
\end{align}
where the coefficients are given by
\begin{equation}
p=1+c_{4}m_{1}^{(4)}+c_{5}m_{1}^{(5)}+\left(d_{17}+d_{18}\right)(m_{1}^{(4)})^{2}+d_{19}m_{1}^{(4)}m_{1}^{(5)}+d_{20}(m_{1}^{(5)})^{2},
\end{equation}
\begin{eqnarray}
q & = & -1+c_{1}m^{(1)}+c_{2}m_{1}^{(4)}+\left(3c_{2}+c_{4}\right)m_{2}^{(4)}+c_{3}m_{1}^{(5)}+\left(3c_{3}+c_{5}\right)m_{2}^{(5)}\nonumber \\
 &  & +d_{1}(m^{(1)})^{2}+\left(6d_{4}+2d_{17}+2d_{18}\right)m_{1}^{(4)}m_{2}^{(4)}+d_{4}(m_{1}^{(4)})^{2}+\left(9d_{4}+3d_{17}+3d_{18}\right)(m_{2}^{(4)})^{2}\nonumber \\
 &  & +d_{5}m_{1}^{(4)}m_{1}^{(5)}+\left(3d_{5}+d_{19}\right)m_{1}^{(4)}m_{2}^{(5)}+\left(3d_{5}+d_{19}\right)m_{2}^{(4)}m_{1}^{(5)}+\left(9d_{5}+3d_{19}\right)m_{2}^{(4)}m_{2}^{(5)}\nonumber \\
 &  & +d_{2}m^{(1)}m_{1}^{(4)}+3d_{2}m^{(1)}m_{2}^{(4)}+\left(6d_{6}+2d_{20}\right)m_{1}^{(5)}m_{2}^{(5)}+(d_{6})(m_{1}^{(5)})^{2}+\left(9d_{6}+3d_{20}\right)(m_{2}^{(5)})^{2}\nonumber \\
 &  & +d_{3}m^{(1)}m_{1}^{(5)}+3d_{3}m^{(1)}m_{2}^{(5)},
\end{eqnarray}
and
\begin{eqnarray}
r & = & (m^{(3)}){}^{2}d_{11}+9(m_{1}^{(6)}){}^{2}d_{14}+6m_{1}^{(6)}m_{2}^{(6)}d_{14}+(m_{2}^{(6)}){}^{2}d_{14}+3(m_{1}^{(6)}){}^{2}d_{15}+7m_{1}^{(6)}m_{2}^{(6)}d_{15}\nonumber \\
 &  & +2(m_{2}^{(6)}){}^{2}d_{15}+(m_{1}^{(6)}){}^{2}d_{16}+4m_{1}^{(6)}m_{2}^{(6)}d_{16}+4(m_{2}^{(6)}){}^{2}d_{16}+3(m_{1}^{(6)}){}^{2}d_{21}+2m_{1}^{(6)}m_{2}^{(6)}d_{21}\nonumber \\
 &  & +2(m_{2}^{(6)}){}^{2}d_{21}+(m_{1}^{(6)}){}^{2}d_{22}+4m_{1}^{(6)}m_{2}^{(6)}d_{22}+\frac{3}{2}(m_{2}^{(6)}){}^{2}d_{22}+(m^{(2)}){}^{2}d_{7}\nonumber \\
 &  & +m^{(2)}\left[2m_{2}^{(6)}d_{10}+m^{(3)}d_{8}+m_{2}^{(6)}d_{9}+m_{1}^{(6)}(d_{10}+3d_{9})+f_{1}\right]\nonumber \\
 &  & +m^{(3)}\left(3m_{1}^{(6)}d_{12}+m_{2}^{(6)}d_{12}+m_{1}^{(6)}d_{13}+2m_{2}^{(6)}d_{13}+f_{2}\right)\nonumber \\
 &  & +3m_{1}^{(6)}f_{3}+m_{2}^{(6)}f_{3}+m_{1}^{(6)}f_{4}+2m_{2}^{(6)}f_{4}.
\end{eqnarray}
The coefficients $p,q,r$ imply that even starting from standard
general relativity, the existence of an independent affine connection
(in terms of the disformation tensor) would inevitably make the effective
Lagrangian for the metric variables take the form of a general SCG theory.

\section{Conclusion} \label{sec:con}

The spatially covariant gravity (SCG), which corresponds to the generally
covariant scalar-tensor theory in the unitary gauge, has been proved
useful in analyzing and extending the original covariant theory. In
this work, we generalize the SCG method to the scalar-nonmetricity
theory, in which the covariant derivatives are not metric compatible
while still torsionless.

The starting point is the 3+1 decomposition of 4-dimensional generally
covariant quantities. In Sec. \ref{sec:pre}, we discuss the 3+1
decomposition of the disformation tensor, the curvature tensor, and
especially the covariant derivatives of the scalar field up to the
third order, given in (\ref{cd1sca_dec}), (\ref{cd2sca_dec})
and (\ref{cd3sca_dec}). The results show explicitly how spatially
covariant quantities arise after taking the unitary gauge. Following
the SCG method, since all the quantities after the 3+1 decomposition are
spatial tensors, we may use them as basic building blocks to construct
the SCG Lagrangian. In particular, we find that the main distinction
of a non-Riemannian theory from a Riemannian one originates from couplings with the disformation tensor $L_{abc}$. As a result, we treat the
projections of the disformation tensor (\ref{projdis}) as basic building
blocks in constructing the Lagrangian (\ref{eq:3.10}). In Sec. \ref{subsec:unig},
we discuss the unitary gauge. We also derive the decomposition
of the Horndeski Lagrangian $\mathcal{L}_{4}^{\text{H}}$, in which
the covariant derivative is not metric compatible. The resulting Lagrangian
in (\ref{L4Hug})-(\ref{L4Hug2}) is nothing but an SCG with nonmetricity,
or more precisely with projections of the disformation tensor. 

In Sec. \ref{sec:mono}, we exhaust and classify all the SCG scalar
monomials up to $d=3$ with $d$ the total number of derivatives.
The results are given in Table \ref{tab:d1}, Table \ref{tab:d2}
and Table \ref{tab:d3}. Up to $d=3$, the affine connection and thus
the nonmetricity tensor act as auxiliary variables. We also follow
the original construction of SCG and assume a nondynamical lapse
function. Therefore, the resulting theory is guaranteed to be ghostfree
and propagate only one scalar degree of freedom.

In Sec. \ref{sec:quad}, we considered the Lagrangian with $d=2$
(\ref{Lagd2}) as an example, which is a polynomial built of linear
combinations of all the scalar monomials of $d=2$. The projections
of the disformation tensor are auxiliary variables, which can be solved
in terms of the metric variables and result in an effective Lagrangian
for the metric variables only. The resulting Lagrangian (\ref{Lageff})
is nothing but an SCG theory of $d=2$ in the Riemannian case, with
modified coefficients.

There are several possible extensions of the results presented in this work. First, in Sec. \ref{sec:quad}  we consider only monomials up to $d=2$, in which the disformation tensor arises without derivatives. Generally, e.g., in the case of $d\geq 3$, there are spatial derivatives acting on the disformation tensor, which not only cause difficulties in solving the disformation tensor but also may introduce nonlocal operators on the metric variables. Second, in this work, we concentrated on the case with nonmetricity, while it is interesting to consider the SCG with a general affine connection following the same procedure in this work.
Third, following \cite{Gao:2020qxy,Gao:2020yzr}, it is interesting to derive the generally covariant correspondence  of the SCG Lagrangians constructed in this work. One may expect to get more general scalar-nonmetricity theories with a single scalar degree of freedom.
Finally, as we have mentioned at the end of Sec. \ref{sec:mono}, the Lie derivatives of the disformation tensor or more general affine connection can be introduced in principle. It is interesting to explore such a possibility and construct consistent theories that are well-behaved and propagate the desired degrees of freedom. We would like to investigate these extensions in the future.

\acknowledgments

This work was partly supported by the National Natural Science Foundation of China (NSFC) under the grant No. 11975020.


\appendix

\section{Explicit expressions for the coefficients in the decomposition of
$\nabla_{c}\nabla_{a}\nabla_{b}\phi$} \label{app:cd3phi}

In this appendix we show the explicit expressions for the coefficients
in decomposition of the third-order derivative of the scalar field
$\nabla_{c}\nabla_{a}\nabla_{b}\phi$ (\ref{cd3sca_dec}). We have
\begin{eqnarray}
U & = & \pounds_{\bm{n}}^{3}\phi-\pounds_{\bm{n}}(a_{d}\text{D}^{d}\phi)+\pounds_{\bm{n}}L^{(1)}\pounds_{\bm{n}}\phi-\pounds_{\bm{n}}(L_{d}^{(3)}\text{D}^{d}\phi)\nonumber \\
 &  & +3L^{(1)}\pounds_{\bm{n}}^{2}\phi-2L^{(1)}a_{d}\text{D}^{d}\phi+2L^{(1)}L^{(1)}\pounds_{\bm{n}}\phi-2L^{(1)}L_{d}^{(3)}\text{D}^{d}\phi\nonumber \\
 &  & +2a^{c}a_{c}\pounds_{\bm{n}}\phi-2a^{c}\pounds_{\bm{n}}\text{D}_{c}\phi+2a^{c}K_{c}^{d}\text{D}_{d}\phi+2a^{c}(L_{dc}^{(4)}\text{D}^{d}\phi-L_{c}^{(2)}\pounds_{\bm{n}}\phi)\nonumber \\
 &  & -2L^{(3)c}(-a_{c}\pounds_{\bm{n}}\phi+\pounds_{\bm{n}}\text{D}_{c}\phi-K_{c}^{d}\text{D}_{d}\phi-L_{dc}^{(4)}\text{D}^{d}\phi+L_{c}^{(2)}\pounds_{\bm{n}}\phi),
\end{eqnarray}
\begin{eqnarray}
V_{b} & = & -2a_{b}\pounds_{\bm{n}}^{2}\phi-(\pounds_{\bm{n}}a_{b}-2K_{b}^{\ c}a_{c})\pounds_{\bm{n}}\phi+\pounds_{\bm{n}}^{2}\text{D}_{b}\phi-\pounds_{\bm{n}}K_{b}^{d}\text{D}_{d}\phi\nonumber \\
 &  & -2K_{b}^{d}\pounds_{\bm{n}}\text{D}_{d}\phi+a_{b}a_{d}\text{D}^{d}\phi+K_{b}^{\ c}K_{c}^{d}\text{D}_{d}\phi-a^{c}\text{D}_{c}\text{D}_{b}\phi-2a_{b}L^{(1)}\pounds_{\bm{n}}\phi\nonumber \\
 &  & -L^{(1)}K_{b}^{d}\text{D}_{d}\phi+2L^{(1)}L_{b}^{(2)}\pounds_{\bm{n}}\phi+L^{(1)}\pounds_{\bm{n}}\text{D}_{b}\phi-L^{(1)}L_{db}^{(4)}\text{D}^{d}\phi-L_{b}^{(2)}L_{d}^{(3)}\text{D}^{d}\phi\nonumber \\
 &  & -K_{b}^{\ c}L_{c}^{(2)}\pounds_{\bm{n}}\phi+\pounds_{\bm{n}}L_{b}^{(2)}\pounds_{\bm{n}}\phi+2L_{b}^{(2)}\pounds_{\bm{n}}^{2}\phi-L_{b}^{(2)}a_{d}\text{D}^{d}\phi-L^{(2)c}L_{cb}^{(4)}\pounds_{\bm{n}}\phi\nonumber \\
 &  & +a_{b}L_{d}^{(3)}\text{D}^{d}\phi+L_{c}^{(3)}K_{b}^{c}\pounds_{\bm{n}}\phi-L_{c}^{(3)}\text{D}^{c}\text{D}_{b}\phi-L_{c}^{(3)}L_{\ \ \ b}^{(5)\ c}\pounds_{\bm{n}}\phi+K_{b}^{\ c}L_{dc}^{(4)}\text{D}^{d}\phi\nonumber \\
 &  & -L_{db}^{(4)}\pounds_{\bm{n}}\text{D}^{d}\phi-\pounds_{\bm{n}}L_{db}^{(4)}\text{D}^{d}\phi+L_{cb}^{(4)}a^{c}\pounds_{\bm{n}}\phi-L_{cb}^{(4)}\pounds_{\bm{n}}\text{D}^{c}\phi+L_{cb}^{(4)}K^{dc}\text{D}_{d}\phi\nonumber \\
 &  & +L_{cb}^{(4)}L_{\ \ \ d}^{(4)\ c}\text{D}^{d}\phi-a^{c}L_{cb}^{(5)}\pounds_{\bm{n}}\phi,
\end{eqnarray}
\begin{align}
W_{c} & =-2a_{c}\pounds_{\bm{n}}^{2}\phi-(\pounds_{\bm{n}}a_{c}-2K_{c}^{\ d}a_{d})\pounds_{\bm{n}}\phi+\pounds_{\bm{n}}^{2}\text{D}_{c}\phi-2K_{c}^{\ d}\pounds_{\bm{n}}\text{D}_{d}\phi-\text{D}_{c}(a_{d}\text{D}^{d}\phi)\nonumber \\
 & \ \ \ \ +2K_{c}^{\ d}K_{d}^{a}\text{D}_{a}\phi+\text{D}_{c}(L^{(1)}\pounds_{\bm{n}}\phi)+2L^{(1)}L_{c}^{(2)}\pounds_{\bm{n}}\phi-2K_{c}^{\ d}L_{d}^{(2)}\pounds_{\bm{n}}\phi+2L_{c}^{(2)}\pounds_{\bm{n}}^{2}\phi\nonumber \\
 & \ \ \ \ -2L_{c}^{(2)}a_{a}\text{D}^{a}\phi-L_{c}^{(2)}L_{a}^{(3)}\text{D}^{a}\phi-2L_{dc}^{(4)}L^{(2)d}\pounds_{\bm{n}}\phi-\text{D}_{c}(L_{a}^{(3)}\text{D}^{a}\phi)+2K_{c}^{\ d}L_{ad}^{(4)}\text{D}^{a}\phi\nonumber \\
 & \ \ \ \ +2L_{dc}^{(4)}a^{d}\pounds_{\bm{n}}\phi-2L_{dc}^{(4)}\pounds_{\bm{n}}\text{D}^{d}\phi+2L_{dc}^{(4)}K^{ad}\text{D}_{a}\phi+2L_{dc}^{(4)}L_{a}^{(4)\ d}\text{D}^{a}\phi,
\end{align}
\begin{align}
X_{ab} & =-K_{ab}\pounds_{\bm{n}}^{2}\phi+(2a_{a}a_{b}+2K_{ac}K_{b}^{c}-\pounds_{\bm{n}}K_{ab})\pounds_{\bm{n}}\phi-2a_{(a}\pounds_{\bm{n}}\text{D}_{b)}\phi+\pounds_{\bm{n}}\text{D}_{a}\text{D}_{b}\phi\nonumber \\
 & \ \ \ \ +2a_{(a}K_{b)}^{c}\text{D}_{c}\phi-2K_{c(a}\text{D}_{b)}\text{D}^{c}\phi-2a_{(a}L_{b)}^{(2)}\pounds_{\bm{n}}\phi-2L_{(a}^{(2)}a_{b)}\pounds_{\bm{n}}\phi+2L_{(a}^{(2)}\pounds_{\bm{n}}\text{D}_{b)}\phi\nonumber \\
 & \ \ \ \ -2L_{(a}^{(2)}K_{b)}^{c}\text{D}_{c}\phi+2L_{a}^{(2)}L_{b}^{(2)}\pounds_{\bm{n}}\phi-2L_{c(a}^{(4)}\text{D}_{b)}\text{D}^{c}\phi-2L_{c(b}^{(4)}L_{a)}^{(2)}\text{D}^{c}\phi\nonumber \\
 & \ \ \ \ +2L_{c(a}^{(4)}K_{b)}^{c}\pounds_{\bm{n}}\phi+2L_{c(b}^{(4)}a_{a)}\text{D}^{c}\phi-2L_{c(a}^{(4)}L_{\ \ \ b)}^{(5)\ c}\pounds_{\bm{n}}\phi-2K_{(ac}L_{\ \ \ b)}^{(5)\ c}\pounds_{\bm{n}}\phi\nonumber \\
 & \ \ \ \ +\pounds_{\bm{n}}L_{ab}^{(5)}\pounds_{\bm{n}}\phi+L_{ab}^{(5)}\pounds_{\bm{n}}^{2}\phi,
\end{align}
\begin{align}
Y_{ca} & =-K_{ca}\pounds_{\bm{n}}^{2}\phi+(K_{cd}K_{a}^{d}-\text{D}_{c}a_{a}+a_{a}a_{c})\pounds_{\bm{n}}\phi-a_{a}\pounds_{\bm{n}}\text{D}_{c}\phi+\text{D}_{c}(\pounds_{\bm{n}}\text{D}_{a}\phi)\nonumber \\
 & \ \ \ \ -\text{D}_{c}K_{a}^{b}\text{D}_{b}\phi+K_{ca}a_{b}\text{D}^{b}\phi-2K_{(a}^{b}\text{D}_{c)}\text{D}_{b}\phi-L^{(1)}K_{ca}\pounds_{\bm{n}}\phi+\text{D}_{c}(L_{a}^{(2)}\pounds_{\bm{n}}\phi)\nonumber \\
 & \ \ \ \ -L_{c}^{(2)}a_{a}\pounds_{\bm{n}}\phi+L_{c}^{(2)}\pounds_{\bm{n}}\text{D}_{a}\phi-L_{c}^{(2)}K_{a}^{b}\text{D}_{b}\phi-L_{c}^{(2)}L_{ba}^{(4)}\text{D}^{b}\phi+L_{c}^{(2)}L_{a}^{(2)}\pounds_{\bm{n}}\phi\nonumber \\
 & \ \ \ \ +K_{ca}L_{b}^{(3)}\text{D}^{b}\phi-\text{D}_{c}(L_{ba}^{(4)}\text{D}^{b}\phi)+L_{dc}^{(4)}K_{a}^{d}\pounds_{\bm{n}}\phi-L_{dc}^{(4)}\text{D}_{a}\text{D}^{d}\phi-L_{dc}^{(4)}L_{\ \ \ a}^{(5)\ d}\pounds_{\bm{n}}\phi\nonumber \\
 & \ \ \ \ -K_{cd}L_{\ \ \ a}^{(5)\ d}\pounds_{\bm{n}}\phi+L_{ac}^{(5)}L^{(1)}\pounds_{\bm{n}}\phi-L_{ac}^{(5)}L_{b}^{(3)}\text{D}^{b}\phi+L_{ac}^{(5)}\pounds_{\bm{n}}^{2}\phi-L_{ac}^{(5)}a_{b}\text{D}^{b}\phi,
\end{align}
and
\begin{align}
Z_{cab} & =(-\text{D}_{c}K_{ab}+3K_{(ab}a_{c)})\pounds_{\bm{n}}\phi-3K_{(ab}\pounds_{\bm{n}}\text{D}_{c)}\phi+2K_{c(a}K_{b)}^{d}\text{D}_{d}\phi+\text{D}_{c}\text{D}_{a}\text{D}_{b}\phi\nonumber \\
 & \ \ \ \ -2K_{c(a}L_{b)}^{(2)}\pounds_{\bm{n}}\phi+2L_{d(b}^{(4)}K_{a)c}\text{D}^{d}\phi-2L_{d(b}^{(4)}L_{ca)}^{(5)}\text{D}^{d}\phi+\text{D}_{c}(L_{ab}^{(5)}\pounds_{\bm{n}}\phi)\nonumber \\
 & \ \ \ \ -2L_{c(a}^{(5)}a_{b)}\pounds_{\bm{n}}\phi+2L_{c(a}^{(5)}\pounds_{\bm{n}}\text{D}_{b)}\phi-2L_{c(a}^{(5)}K_{b)}^{d}\text{D}_{d}\phi+2L_{c(a}^{(5)}L_{b)}^{(2)}\pounds_{\bm{n}}\phi.
\end{align}

\section{Relations and decomposition of the curvature tensor} \label{app:deccurv}

In this appendix we show the relation between the spacetime curvature
tensor $R_{\phantom{a}bcd}^{a}$ and the curvature tensor adapted
to the Levi-Civita connection $\mathring{R}_{\phantom{a}bcd}^{a}$
as well as their decomposition. The Riemann tensors $R_{\phantom{a}bcd}^{a}$
and $\mathring{R}_{\phantom{a}bcd}^{a}$ are related by
\begin{align}
R_{\phantom{a}bcd}^{a}-\mathring{R}_{\phantom{a}bcd}^{a} & =\mathring{\nabla}_{c}L_{\phantom{a}bd}^{a}-\mathring{\nabla}_{d}L_{\phantom{a}bc}^{a}+L_{\phantom{a}ec}^{a}L_{\phantom{e}bd}^{e}-L_{\phantom{a}ed}^{a}L_{\phantom{e}bc}^{e}\nonumber \\
 & =\nabla_{c}L_{\phantom{a}bd}^{a}-\nabla_{d}L_{\phantom{a}bc}^{a}+L_{\phantom{e}bc}^{e}L_{\phantom{a}ed}^{a}-L_{\phantom{e}bd}^{e}L_{\phantom{a}ec}^{a}.\label{app:RieTen_rel}
\end{align}
By taking into account of the antisymmetry of the last two indices
of the curvature tensors, there will be 8 independent decomposition
of (\ref{app:RieTen_rel}), which are given by
\begin{eqnarray}
 &  & R_{\bm{n}\bm{n}\bm{n}\hat{d}}-\mathring{R}_{\bm{n}\bm{n}\bm{n}\hat{d}}\nonumber \\
 & = & -a_{d}L^{(1)}+K^{e}{}_{d}L^{(2)}{}_{e}+K^{e}{}_{d}L^{(3)}{}_{e}-a^{e}L^{(4)}{}_{ed}-L^{(3)}{}^{e}L^{(4)}{}_{ed}+L^{(2)}{}_{e}L^{(4)}{}^{e}{}_{d}\nonumber \\
 &  & +L^{(3)}{}_{e}L^{(4)}{}^{e}{}_{d}-a^{e}L^{(5)}{}_{de}-L^{(3)}{}^{e}L^{(5)}{}_{de}-\text{D}_{d}L^{(1)}+\pounds_{\bm{n}}L^{(2)}{}_{d},\label{eq:B.4}
\end{eqnarray}
\begin{eqnarray}
 &  & R_{\bm{n}\bm{n}\hat{c}\hat{d}}-\mathring{R}_{\bm{n}\bm{n}\hat{c}\hat{d}}\nonumber \\
 & = & K_{cd}L^{(1)}-K_{dc}L^{(1)}+K^{e}{}_{d}L^{(4)}{}_{ec}-K^{e}{}_{c}L^{(4)}{}_{ed}-2L^{(4)}{}_{ed}L^{(4)}{}^{e}{}_{c}+2L^{(4)}{}_{ec}L^{(4)}{}^{e}{}_{d}\nonumber \\
 &  & +K^{e}{}_{d}L^{(5)}{}_{ce}+L^{(4)}{}^{e}{}_{d}L^{(5)}{}_{ce}-K^{e}{}_{c}L^{(5)}{}_{de}-L^{(4)}{}^{e}{}_{c}L^{(5)}{}_{de}+L^{(2)}{}^{e}L^{(6)}{}_{ecd},\nonumber \\
 &  & -L^{(2)}{}^{e}L^{(6)}{}_{edc}+\text{D}_{c}L^{(2)}{}_{d}-\text{D}_{d}L^{(2)}{}_{c},
\end{eqnarray}
\begin{eqnarray}
 &  & R_{\bm{n}\hat{b}\bm{n}\hat{d}}-\mathring{R}_{\bm{n}\hat{b}\bm{n}\hat{d}}\nonumber \\
 & = & K_{bd}L^{(1)}-a_{d}L^{(2)}{}_{b}-a_{b}L^{(2)}{}_{d}+L^{(2)}{}_{b}L^{(2)}{}_{d}+K^{e}{}_{d}L^{(4)}{}_{eb}-L^{(4)}{}_{ed}L^{(4)}{}^{e}{}_{b}+L^{(4)}{}_{eb}L^{(4)}{}^{e}{}_{d}\nonumber \\
 &  & -L^{(1)}L^{(5)}{}_{db}-K^{e}{}_{b}L^{(5)}{}_{de}-L^{(4)}{}^{e}{}_{b}L^{(5)}{}_{de}-a^{e}L^{(6)}{}_{edb}-\text{D}_{d}L^{(2)}{}_{b}+\pounds_{\bm{n}}L^{(5)}{}_{bd},
\end{eqnarray}
\begin{eqnarray}
 &  & R_{\bm{n}\hat{b}\hat{c}\hat{d}}-\mathring{R}_{\bm{n}\hat{b}\hat{c}\hat{d}}\nonumber \\
 & = & K_{cd}L^{(2)}{}_{b}-K_{dc}L^{(2)}{}_{b}+K_{bd}L^{(2)}{}_{c}-K_{bc}L^{(2)}{}_{d}+L^{(2)}{}_{d}L^{(5)}{}_{cb}-L^{(2)}{}_{c}L^{(5)}{}_{db}+K^{e}{}_{d}L^{(6)}{}_{ecb}\nonumber \\
 &  & +L^{(5)}{}^{e}{}_{b}L^{(6)}{}_{ecd}-K^{e}{}_{c}L^{(6)}{}_{edb}-L^{(5)}{}^{e}{}_{b}L^{(6)}{}_{edc}+\text{D}_{c}L^{(5)}{}_{db}-\text{D}_{d}L^{(5)}{}_{cb},
\end{eqnarray}
\begin{eqnarray}
 &  & R_{\hat{a}\bm{n}\bm{n}\hat{d}}-\mathring{R}_{\hat{a}\bm{n}\bm{n}\hat{d}}\nonumber \\
 & = & K_{ad}L^{(1)}-a_{a}L^{(2)}{}_{d}-a_{d}L^{(3)}{}_{a}-L^{(2)}{}_{d}L^{(3)}{}_{a}+L^{(1)}L^{(4)}{}_{ad}+K^{e}{}_{d}L^{(4)}{}_{ae}-K^{e}{}_{a}L^{(4)}{}_{ed}\nonumber \\
 &  & -L^{(4)}{}_{ed}L^{(4)}{}^{e}{}_{a}+L^{(4)}{}_{ae}L^{(4)}{}^{e}{}_{d}+L^{(4)}{}_{ea}L^{(4)}{}^{e}{}_{d}+L^{(1)}L^{(5)}{}_{ad}-L^{(1)}L^{(5)}{}_{da}-a^{e}L^{(6)}{}_{ade}\nonumber \\
 &  & -L^{(3)}{}^{e}L^{(6)}{}_{ade}-L^{(3)}{}^{e}L^{(6)}{}_{ead}-\text{D}_{d}L^{(3)}{}_{a}+\pounds_{\bm{n}}L^{(4)}{}_{ad},
\end{eqnarray}
\begin{eqnarray}
 &  & R_{\hat{a}\bm{n}\hat{c}\hat{d}}-\mathring{R}_{\hat{a}\bm{n}\hat{c}\hat{d}}\nonumber \\
 & = & K_{ad}L^{(2)}{}_{c}-K_{ac}L^{(2)}{}_{d}+K_{cd}L^{(3)}{}_{a}-K_{dc}L^{(3)}{}_{a}-L^{(2)}{}_{d}L^{(4)}{}_{ac}+L^{(2)}{}_{c}L^{(4)}{}_{ad}-L^{(2)}{}_{d}L^{(5)}{}_{ac}\nonumber \\
 &  & +L^{(2)}{}_{c}L^{(5)}{}_{ad}+L^{(2)}{}_{d}L^{(5)}{}_{ca}-L^{(2)}{}_{c}L^{(5)}{}_{da}+K^{e}{}_{d}L^{(6)}{}_{ace}+L^{(4)}{}^{e}{}_{d}L^{(6)}{}_{ace}-K^{e}{}_{c}L^{(6)}{}_{ade}\nonumber \\
 &  & -L^{(4)}{}^{e}{}_{c}L^{(6)}{}_{ade}+L^{(4)}{}^{e}{}_{d}L^{(6)}{}_{eac}-L^{(4)}{}^{e}{}_{c}L^{(6)}{}_{ead}+L^{(4)}{}_{a}{}^{e}L^{(6)}{}_{ecd}-L^{(4)}{}_{a}{}^{e}L^{(6)}{}_{edc}\nonumber \\
 &  & +\text{D}_{c}L^{(4)}{}_{ad}-\text{D}_{d}L^{(4)}{}_{ac},
\end{eqnarray}
\begin{eqnarray}
 &  & R_{\hat{a}\hat{b}\bm{n}\hat{d}}-\mathring{R}_{\hat{a}\hat{b}\bm{n}\hat{d}}\nonumber \\
 & = & K_{ad}L^{(2)}{}_{b}+K_{bd}L^{(3)}{}_{a}-a_{d}L^{(4)}{}_{ab}-a_{b}L^{(4)}{}_{ad}+L^{(2)}{}_{b}L^{(4)}{}_{ad}+L^{(2)}{}_{b}L^{(5)}{}_{ad}-L^{(2)}{}_{b}L^{(5)}{}_{da}\nonumber \\
 &  & -a_{a}L^{(5)}{}_{db}-L^{(3)}{}_{a}L^{(5)}{}_{db}-K^{e}{}_{b}L^{(6)}{}_{ade}-L^{(4)}{}^{e}{}_{b}L^{(6)}{}_{ade}-L^{(4)}{}^{e}{}_{b}L^{(6)}{}_{ead}-K^{e}{}_{a}L^{(6)}{}_{edb}\nonumber \\
 &  & -\text{D}_{d}L^{(4)}{}_{ab}+\pounds_{\bm{n}}L^{(6)}{}_{abd},
\end{eqnarray}
and
\begin{eqnarray}
 &  & R_{\hat{a}\hat{b}\hat{c}\hat{d}}-\mathring{R}_{\hat{a}\hat{b}\hat{c}\hat{d}}\nonumber \\
 & = & K_{cd}L^{(4)}{}_{ab}-K_{dc}L^{(4)}{}_{ab}+K_{bd}L^{(4)}{}_{ac}-K_{bc}L^{(4)}{}_{ad}+K_{ad}L^{(5)}{}_{cb}+L^{(4)}{}_{ad}L^{(5)}{}_{cb}+L^{(5)}{}_{ad}L^{(5)}{}_{cb}\nonumber \\
 &  & -L^{(5)}{}_{cb}L^{(5)}{}_{da}-K_{ac}L^{(5)}{}_{db}-L^{(4)}{}_{ac}L^{(5)}{}_{db}-L^{(5)}{}_{ac}L^{(5)}{}_{db}+L^{(5)}{}_{ca}L^{(5)}{}_{db}-L^{(6)}{}_{ead}L^{(6)}{}^{e}{}_{cb}\nonumber \\
 &  & +L^{(6)}{}_{abe}L^{(6)}{}^{e}{}_{cd}+L^{(6)}{}_{eac}L^{(6)}{}^{e}{}_{db}-L^{(6)}{}_{abe}L^{(6)}{}^{e}{}_{dc}+\text{D}_{c}L^{(6)}{}_{adb}-\text{D}_{d}L^{(6)}{}_{acb}.\label{eq:B.11}
\end{eqnarray}
Keep in mind that in the above the disformation tensor is also included
implicitly in the spatial covariant derivative $\mathrm{D}_{a}$.

By contracting indices of (\ref{app:RieTen_rel}), we get the relation
between Ricci tensors
\begin{align}
R_{bd}-\mathring{R}_{bd} & =\nabla_{a}L_{\phantom{a}bd}^{a}-\nabla_{d}L_{\phantom{a}ba}^{a}+L_{\phantom{e}ba}^{e}L_{\phantom{a}ed}^{a}-L_{\phantom{e}bd}^{e}L_{\phantom{a}ea}^{a}.
\end{align}
Note that $R_{ab}$ is not symmetric. The above relation implies that
the spatial Ricci tensors $^{3}\!R_{ab}$ and $^{3}\!\mathring{R}_{ab}$
(defined in (\ref{RicTen3d}) and (\ref{RicTen3d_LC}), respectively)
are not independent
\begin{eqnarray}
^{3}\!R_{ab} & = & h_{e}^{c}h_{a}^{f}h_{b}^{d}R_{\phantom{e}fcd}^{e}+2K_{a[c}K_{b]}^{c}\nonumber \\
 & = & ^{3}\!\mathring{R}_{ab}+\text{D}^{c}L^{(6)}{}_{cab}-\text{D}_{b}L^{(6)}{}_{ca}^{\ \ c}+L_{\ \ \ \ c}^{(6)e\ c}L^{(6)}{}_{eab}-L_{\ \ \ \ cb}^{(6)e}L^{(6)}{}_{ea}^{\ \ c}\nonumber \\
 &  & -L_{ab}^{(5)}K-L_{cb}^{(4)}K_{a}^{c}+L^{(5)}{}_{ac}K_{b}^{c}+L^{(4)}{}_{c}^{\ c}K_{ba}-L^{(4)}{}_{c}^{\phantom{c}c}L^{(5)}{}_{ab}+L^{(4)}{}_{cb}L^{(5)}{}_{a}^{\phantom{a}c},\label{eq:B.3}
\end{eqnarray}
which are related to each other by the disformation tensor and its
spatial derivative. 

Note that in (\ref{eq:B.3}) there is no temporal derivative on the disformation
tensor. Therefore, it is equivalent to use either $^{3}\!R_{ab}$
or $^{3}\!\mathring{R}_{ab}$ as the basic variable (together with
the disformation tensor) to build the Lagrangian. However, if we consider
a more general Lagrangian as a function of $R_{abcd}$, the Lie derivative
of the disformation tensor $\pounds_{\bm{n}}L$ has to be taken into
account, as a result of the decomposition of equation (\ref{app:RieTen_rel}). 

\section{The solution for the disformation tensor} \label{app:soldisten}

Varying the Lagrangian (\ref{Lagd2}) with respect to $L^{(i)}$'s
yields the following equations of motion
\begin{eqnarray}
0 & = & c_{1}K+2d_{1}L^{(1)}+d_{2}L_{\ \ \ i}^{(4)\ i}+d_{3}L_{\ \ \ i}^{(5)\ i},\\
0 & = & 2d_{7}L_{i}^{(2)}+d_{8}L_{i}^{(3)}+d_{9}L_{\ \ \ i\ j}^{(6)\ j}+d_{10}L_{\ \ \ \ ij}^{(6)j}+f_{1}a_{i},\\
0 & = & d_{8}L_{i}^{(2)}+2d_{11}L_{i}^{(3)}+d_{12}L_{\ \ \ i\ j}^{(6)\ j}+d_{13}L_{\ \ \ \ ij}^{(6)j}+f_{2}a_{i},\\
0 & = & c_{2}Kh_{ij}+c_{4}K_{ij}+d_{2}L^{(1)}h_{ij}+2d_{4}L_{\ \ \ k}^{(4)\ k}h_{ij}+d_{5}L_{\ \ \ k}^{(5)\ k}h_{ij}\nonumber \\
 &  & +2d_{17}L_{ij}^{(4)}+2d_{18}L_{ji}^{(4)}+d_{19}L_{ij}^{(5)},\\
0 & = & c_{5}K_{ij}+c_{3}Kh_{ij}+d_{3}L^{(1)}h_{ij}+d_{5}L_{\ \ \ k}^{(4)\ k}h_{ij}+2d_{6}L_{\ \ \ k}^{(5)\ k}h_{ij}+d_{19}L_{(ij)}^{(4)}+2d_{20}L_{ij}^{(5)},\\
0 & = & d_{9}L_{k}^{(2)}h_{ij}+d_{10}L_{(i}^{(2)}h_{j)k}+d_{12}L_{k}^{(3)}h_{ij}+d_{13}L_{(i}^{(3)}h_{j)k}+2d_{14}L_{\ \ \ k\ m}^{(6)\ m}h_{ij}+d_{15}L_{\ \ \ \ km}^{(6)m}h_{ij}\nonumber \\
 &  & +d_{15}h_{k(j}L_{\ \ \ i)\ m}^{(6)\ \ m}+2d_{16}h_{k(j}L_{\ \ \ \ i)m}^{(6)m}+2d_{21}L_{kij}^{(6)}+2d_{22}L_{(ij)k}^{(6)}+f_{3}a_{k}h_{ij}+f_{4}a_{(i}h_{kj)}.\qquad
\end{eqnarray}
For later convenience, we rewrite the above equations to be
\begin{eqnarray}
0 & = & c_{1}K+2d_{1}L^{(1)}+d_{2}h^{ij}L_{ij}^{(4)}+d_{3}h^{ij}L_{ij}^{(5)},\label{eq:C.7}\\
0 & = & 2d_{7}L_{i}^{(2)}+d_{8}L_{i}^{(3)}+\left(d_{9}\delta_{i}^{k}h^{lj}+d_{10}\delta_{i}^{l}h^{kj}\right)L_{klj}^{(6)}+f_{1}a_{i},\label{eq:C.8}\\
0 & = & d_{8}L_{i}^{(2)}+2d_{11}L_{i}^{(3)}+\left(d_{12}\delta_{i}^{k}h^{lj}+d_{13}\delta_{i}^{l}h^{kj}\right)L_{klj}^{(6)}+f_{2}a_{i},\label{eq:C.9}\\
0 & = & c_{2}Kh_{ij}+c_{4}K_{ij}+d_{2}L^{(1)}h_{ij}+\left(2d_{4}h^{kl}h_{ij}+2d_{17}\delta_{i}^{k}\delta_{j}^{l}+2d_{18}\delta_{j}^{k}\delta_{i}^{l}\right)L_{kl}^{(4)}\nonumber \\
 &  & +\left(d_{5}h^{kl}h_{ij}+d_{19}\delta_{i}^{k}\delta_{j}^{l}\right)L_{kl}^{(5)},\label{eq:C.10}\\
0 & = & c_{5}K_{ij}+c_{3}Kh_{ij}+d_{3}L^{(1)}h_{ij}+\left(d_{5}h^{kl}h_{ij}+d_{19}\delta_{(i}^{k}\delta_{j)}^{l}\right)L_{kl}^{(4)}\nonumber \\
 &  & +\left(2d_{6}h^{kl}h_{ij}+2d_{20}\delta_{i}^{k}\delta_{j}^{l}\right)L_{kl}^{(5)},\label{eq:C.11}\\
0 & = & \left(d_{9}h_{ij}\delta_{k}^{l}+d_{10}\delta_{(i}^{l}h_{j)k}\right)L_{l}^{(2)}+\left(d_{12}h_{ij}\delta_{k}^{l}+d_{13}\delta_{(i}^{l}h_{j)k}\right)L_{l}^{(3)}+f_{3}a_{k}h_{ij}+f_{4}a_{(i}h_{kj)}\nonumber \\
 &  & +\Big(2d_{14}h_{ij}\delta_{k}^{l}h^{mn}+d_{15}h_{ij}\delta_{k}^{m}h^{nl}+d_{15}h_{k(j}\delta_{i)}^{l}h^{mn}\nonumber \\
 &  & \qquad+2d_{16}h_{k(j}\delta_{i)}^{m}h^{nl}+2d_{21}\delta_{k}^{l}\delta_{i}^{m}\delta_{j}^{n}+2d_{22}\delta_{(i}^{l}\delta_{j)}^{m}\delta_{k}^{n}\Big)L_{lmn}^{(6)}.\label{eq:C.12}
\end{eqnarray}
 Plugging the ansatz (\ref{eq:4.2})-(\ref{eq:4.7}) into (\ref{eq:C.7})-(\ref{eq:C.12}) yields
	\begin{equation}
		0=\left(c_{1}+2m^{(1)}d_{1}+d_{2}m_{1}^{(4)}+3d_{2}m_{2}^{(4)}+d_{3}m_{1}^{(5)}+3d_{3}m_{2}^{(5)}\right)K,
	\end{equation}
	\begin{equation}
		0=\left[2d_{7}m^{(2)}+d_{8}m^{(3)}+m_{1}^{(6)}\left(3d_{9}+d_{10}\right)+m_{2}^{(6)}\left(d_{9}+2d_{10}\right)+f_{1}\right]a_{i},
	\end{equation}
	\begin{equation}
		0=\left[d_{8}m^{(2)}+2d_{11}m^{(3)}+m_{1}^{(6)}\left(3d_{12}+d_{13}\right)+m_{2}^{(6)}\left(d_{12}+2d_{13}\right)+f_{2}\right]a_{i},
	\end{equation}
	\begin{eqnarray}
		0 & = & \left[c_{4}+m_{1}^{(4)}(2d_{17}+2d_{18})+m_{1}^{(5)}d_{19}\right]K_{ij}+\Big[c_{2}+d_{2}m^{(1)}+2m_{1}^{(4)}d_{4}\nonumber \\
		&  & \qquad+m_{2}^{(4)}\left(6d_{4}+2d_{17}+2d_{18}\right)+m_{1}^{(5)}d_{5}+m_{2}^{(5)}\left(3d_{5}+d_{19}\right)\Big]Kh_{ij},
	\end{eqnarray}
	\begin{eqnarray}
		0 & = & \left(c_{5}+m_{1}^{(4)}d_{19}+2m_{1}^{(5)}d_{20}\right)K_{ij}+\Big[c_{3}+d_{3}m^{(1)}+m_{1}^{(4)}d_{5}\nonumber \\
		&  & \qquad+m_{2}^{(4)}\left(3d_{5}+d_{19}\right)+2m_{1}^{(5)}d_{6}+m_{2}^{(5)}\left(6d_{6}+2d_{20}\right)\Big]Kh_{ij},
	\end{eqnarray}
	\begin{eqnarray}
		0 & = & \left[m^{(2)}d_{9}+m^{(3)}d_{12}+m_{1}^{(6)}\left(6d_{14}+d_{15}+2d_{21}\right)+m_{2}^{(6)}\left(2d_{14}+2d_{15}+d_{22}\right)+f_{3}\right]a_{k}h_{ij}\nonumber \\
		&  & +\Big[m^{(2)}d_{10}+m^{(3)}d_{13}+m_{1}^{(6)}\left(3d_{15}+2d_{16}+2d_{22}\right)\nonumber \\
		&  & \qquad+m_{2}^{(6)}\left(d_{15}+4d_{16}+2d_{21}+d_{22}\right)+f_{4}\Big]a_{(i}h_{j)k}.
	\end{eqnarray}
The above equations are satisfied only if the coefficients of each
monomials (such as $K$, $a_{i}$ etc.) are vanishing. This yields
a set of linear algebraic equations for the coefficients $m^{(1)},m^{(2)},\cdots$,
which can be written as
\begin{equation}
\bm{A}\bm{M}_{1}=\bm{C},
\end{equation}
and
\begin{equation}
\bm{B}\bm{M}_{2}=\bm{F},
\end{equation}
with 
\begin{equation}
\bm{A}=\begin{pmatrix}2d_{1} & d_{2} & 3d_{2} & d_{3} & 3d_{3}\\
d_{2} & 2d_{4} & 6d_{4}+2d_{17}+2d_{18} & d_{5} & 3d_{5}+d_{19}\\
d_{3} & d_{5} & 3d_{5}+d_{19} & d_{6} & 6d_{6}+2d_{20}\\
0 & 2d_{17}+2d_{18} & 0 & d_{19} & 0\\
0 & d_{19} & 0 & d_{20} & 0
\end{pmatrix},
\end{equation}
\begin{equation}
\bm{B}=\begin{pmatrix}2d_{7} & d_{8} & 3d_{9}+d_{10} & d_{9}+2d_{10}\\
d_{8} & 2d_{11} & 3d_{12}+d_{13} & d_{12}+2d_{13}\\
d_{9} & d_{12} & 6d_{14}+d_{15}+2d_{21} & 2d_{14}+2d_{15}+d_{22}\\
d_{10} & d_{13} & 3d_{15}+2d_{16}+2d_{22} & d_{15}+4d_{16}+2d_{21}+d_{22}
\end{pmatrix},
\end{equation}
\begin{eqnarray}
\bm{M}_{1} & = & \left(\begin{array}{ccccc}
m^{(1)} & m_{1}^{(4)} & m_{2}^{(4)} & m_{1}^{(5)} & m_{2}^{(5)}\end{array}\right)^{\mathrm{T}},\\
\bm{C} & = & -\left(\begin{array}{ccccc}
c_{1} & c_{2} & c_{3} & c_{4} & c_{5}\end{array}\right)^{\mathrm{T}},\\
\bm{M}_{2} & = & \left(\begin{array}{cccc}
m^{(2)} & m^{(3)} & m_{1}^{(6)} & m_{2}^{(6)}\end{array}\right)^{\mathrm{T}},\\
\bm{F} & = & -\left(\begin{array}{cccc}
f_{1} & f_{2} & f_{3} & f_{4}\end{array}\right)^{\mathrm{T}}.
\end{eqnarray}

If the matrix $\bm{A}$ and $\bm{B}$ are not degenerated, the solutions
for the coefficients take the form
\begin{align}
m^{(1)} & =\left(\det\bm{A}\right)^{-1}\det\left(\begin{array}{ccccc}
-c_{1} & d_{2} & 3d_{2} & d_{3} & 3d_{3}\\
-c_{2} & 2d_{4} & 6d_{4}+2d_{17}+2d_{18} & d_{5} & 3d_{5}+d_{19}\\
-c_{3} & d_{5} & 3d_{5}+d_{19} & d_{6} & 6d_{6}+2d_{20}\\
-c_{4} & 2d_{17}+2d_{18} & 0 & d_{19} & 0\\
-c_{5} & d_{19} & 0 & d_{20} & 0
\end{array}\right),\\
m_{1}^{(4)} & =\left(\det\bm{A}\right)^{-1}\det\left(\begin{array}{ccccc}
2d_{1} & -c_{1} & 3d_{2} & d_{3} & 3d_{3}\\
d_{2} & -c_{2} & 6d_{4}+2d_{17}+2d_{18} & d_{5} & 3d_{5}+d_{19}\\
d_{3} & -c_{3} & 3d_{5}+d_{19} & d_{6} & 6d_{6}+2d_{20}\\
0 & -c_{4} & 0 & d_{19} & 0\\
0 & -c_{5} & 0 & d_{20} & 0
\end{array}\right),\\
m_{2}^{(4)} & =\left(\det\bm{A}\right)^{-1}\det\left(\begin{array}{ccccc}
2d_{1} & d_{2} & -c_{1} & d_{3} & 3d_{3}\\
d_{2} & 2d_{4} & -c_{2} & d_{5} & 3d_{5}+d_{19}\\
d_{3} & d_{5} & -c_{3} & d_{6} & 6d_{6}+2d_{20}\\
0 & 2d_{17}+2d_{18} & -c_{4} & d_{19} & 0\\
0 & d_{19} & -c_{5} & d_{20} & 0
\end{array}\right),\\
m_{1}^{(5)} & =\left(\det\bm{A}\right)^{-1}\det\left(\begin{array}{ccccc}
2d_{1} & d_{2} & 3d_{2} & -c_{1} & 3d_{3}\\
d_{2} & 2d_{4} & 6d_{4}+2d_{17}+2d_{18} & -c_{2} & 3d_{5}+d_{19}\\
d_{3} & d_{5} & 3d_{5}+d_{19} & -c_{3} & 6d_{6}+2d_{20}\\
0 & 2d_{17}+2d_{18} & 0 & -c_{4} & 0\\
0 & d_{19} & 0 & -c_{5} & 0
\end{array}\right),\\
m_{2}^{(5)} & =\left(\det\bm{A}\right)^{-1}\det\left(\begin{array}{ccccc}
2d_{1} & d_{2} & 3d_{2} & d_{3} & -c_{1}\\
d_{2} & 2d_{4} & 6d_{4}+2d_{17}+2d_{18} & d_{5} & -c_{2}\\
d_{3} & d_{5} & 3d_{5}+d_{19} & d_{6} & -c_{3}\\
0 & 2d_{17}+2d_{18} & 0 & d_{19} & -c_{4}\\
0 & d_{19} & 0 & d_{20} & -c_{5}
\end{array}\right),
\end{align}
and
\begin{align}
m^{(2)} & =\left(\det\bm{B}\right)^{-1}\det\left(\begin{array}{cccc}
-f_{1} & d_{8} & 3d_{9}+d_{10} & d_{9}+2d_{10}\\
-f_{2} & 2d_{11} & 3d_{12}+d_{13} & d_{12}+2d_{13}\\
-f_{3} & d_{12} & 6d_{14}+d_{15}+2d_{21} & 2d_{14}+2d_{15}+d_{22}\\
-f_{4} & d_{13} & 3d_{15}+2d_{16}+2d_{22} & d_{15}+4d_{16}+2d_{21}+d_{22}
\end{array}\right),\\
m^{(3)} & =\left(\det\bm{B}\right)^{-1}\det\left(\begin{array}{cccc}
2d_{7} & -f_{1} & 3d_{9}+d_{10} & d_{9}+2d_{10}\\
d_{8} & -f_{2} & 3d_{12}+d_{13} & d_{12}+2d_{13}\\
d_{9} & -f_{3} & 6d_{14}+d_{15}+2d_{21} & 2d_{14}+2d_{15}+d_{22}\\
d_{10} & -f_{4} & 3d_{15}+2d_{16}+2d_{22} & d_{15}+4d_{16}+2d_{21}+d_{22}
\end{array}\right),\\
m_{1}^{(6)} & =\left(\det\bm{B}\right)^{-1}\det\left(\begin{array}{cccc}
2d_{7} & d_{8} & -f_{1} & d_{9}+2d_{10}\\
d_{8} & 2d_{11} & -f_{2} & d_{12}+2d_{13}\\
d_{9} & d_{12} & -f_{3} & 2d_{14}+2d_{15}+d_{22}\\
d_{10} & d_{13} & -f_{4} & d_{15}+4d_{16}+2d_{21}+d_{22}
\end{array}\right),\\
m_{2}^{(6)} & =\left(\det\bm{B}\right)^{-1}\det\left(\begin{array}{cccc}
2d_{7} & d_{8} & 3d_{9}+d_{10} & -f_{1}\\
d_{8} & 2d_{11} & 3d_{12}+d_{13} & -f_{2}\\
d_{9} & d_{12} & 6d_{14}+d_{15}+2d_{21} & -f_{3}\\
d_{10} & d_{13} & 3d_{15}+2d_{16}+2d_{22} & -f_{4}
\end{array}\right).
\end{align}

\rule[0.5ex]{1\columnwidth}{1pt}

\providecommand{\href}[2]{#2}\begingroup\raggedright\endgroup

\end{document}